\RequirePackage{fix-cm}
\documentclass[smallcondensed]{svjour3}     
\smartqed  
\usepackage{graphicx}
\graphicspath{{figures/}}
\usepackage{natbib}
\usepackage{mathtools}
\usepackage{amssymb,mathrsfs}
\usepackage{multirow}
\usepackage[linkcolor=blue,citecolor=blue,colorlinks]{hyperref}
\usepackage{dsfont}
\DeclareMathAlphabet{\mathpzc}{OT1}{pzc}{m}{it}
\renewcommand{\deg}{^\circ}

\newcommand{\e}{\mathrm{e}}
\newcommand{\ii}{\mathrm{i}}
\newcommand{\Year}{a}
\newcommand{\trans}[1]{{#1}^{\mathrm T}}

\newcommand{\hvec}[1]{\hat{\vec {#1}}}
\newcommand{\mat}[1]{\vec{#1}}
\newcommand{\field}[1]{{\mathrm{#1}}}
\newcommand{\vfield}[1]{{\mathrm{#1}}}
\newcommand{\dd}{\mathrm{d}}
\newcommand{\reduced}[1]{\underbar{#1}}
\newcommand{\inertia}{{\mathds I}}
\newcommand{\invinertia}{{\mathds Q}}

\DeclareMathOperator{\diag}{diag}

\newcommand{\Frac}[2]{\displaystyle\frac{#1}{#2}}
\renewcommand{\equiv}{\coloneqq}
\newcommand{\figmodels}{%
\begin{figure}
\begin{center}
\includegraphics[width=0.33\linewidth]{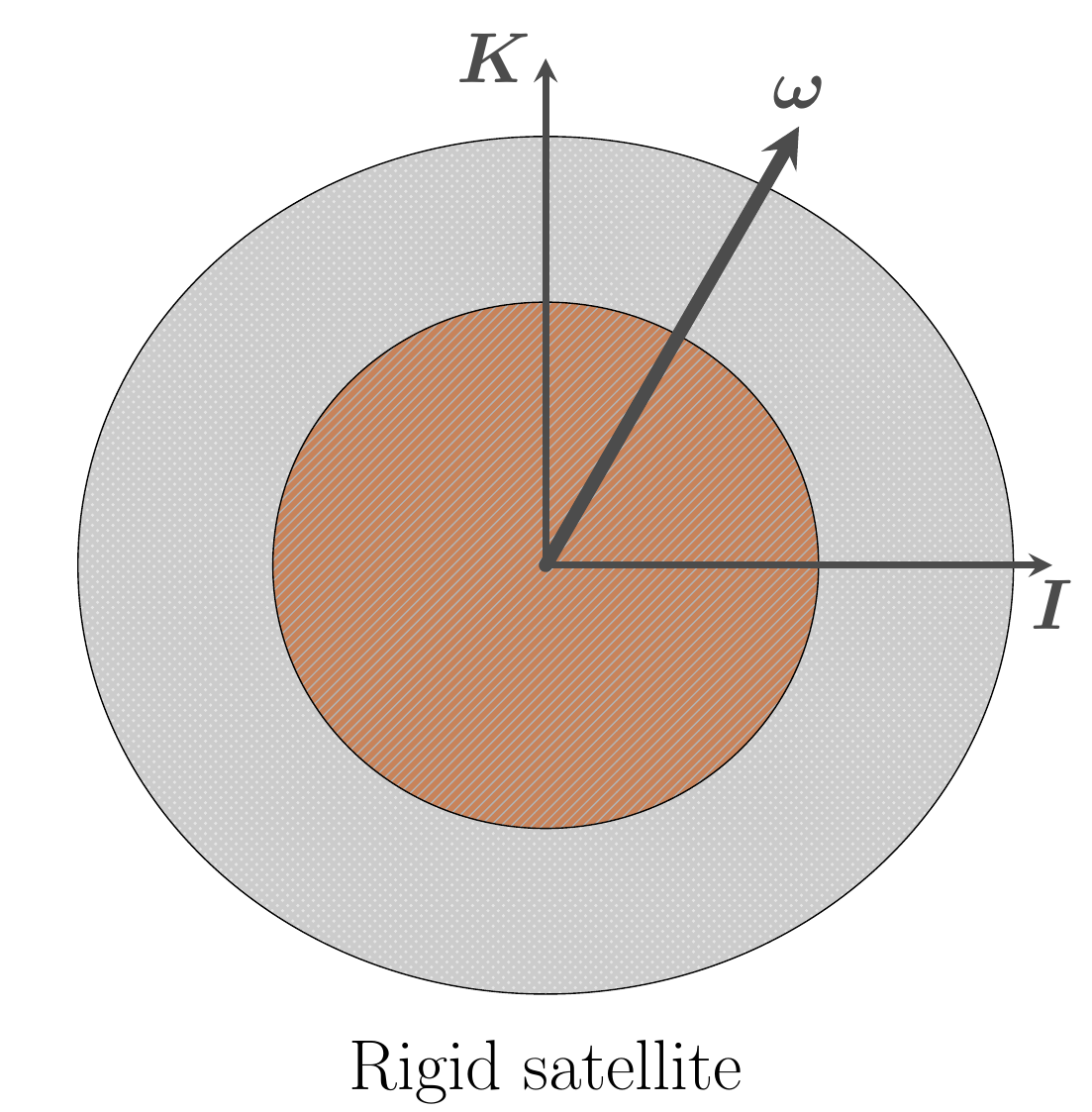}%
\includegraphics[width=0.33\linewidth]{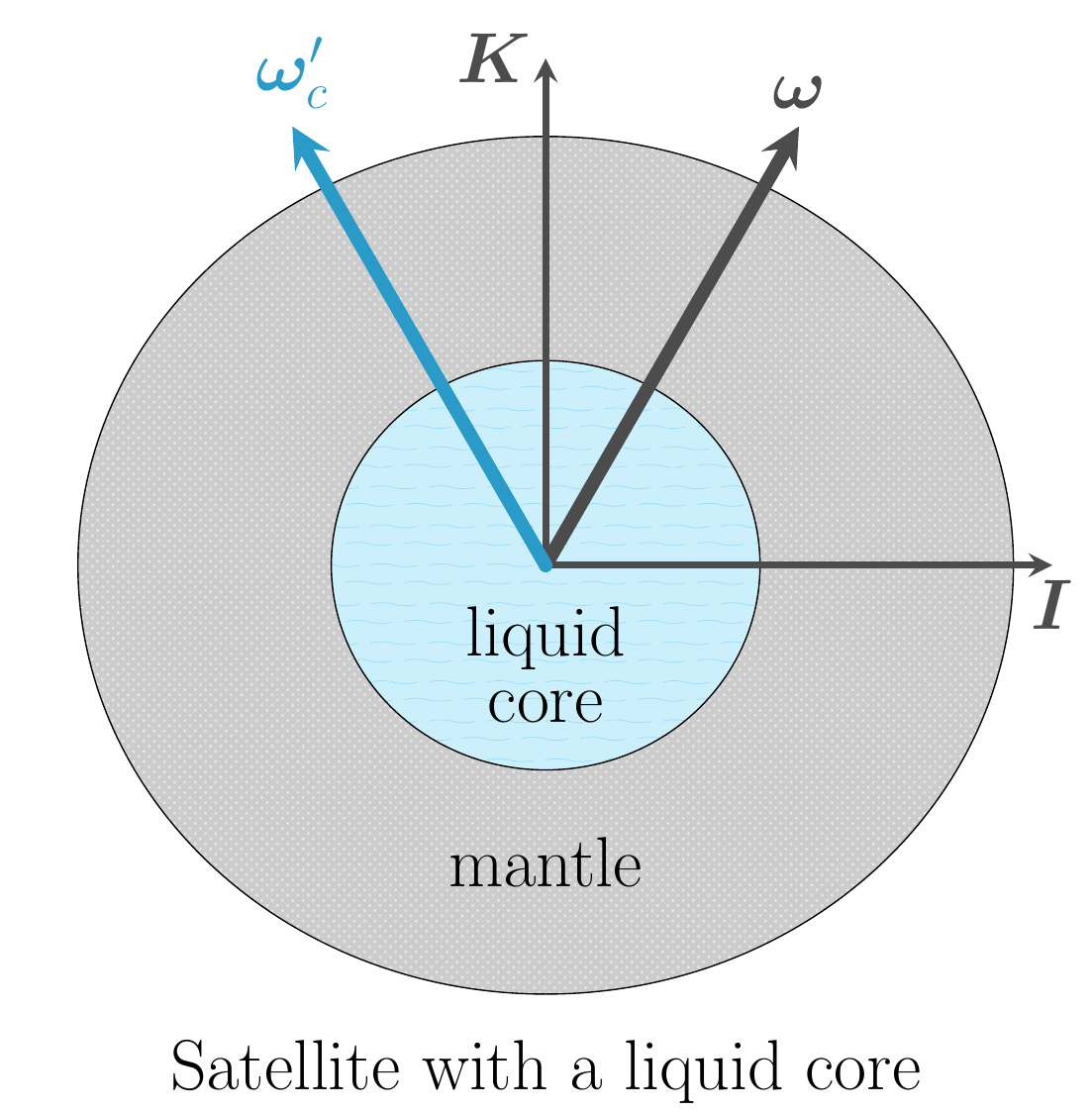}%
\includegraphics[width=0.33\linewidth]{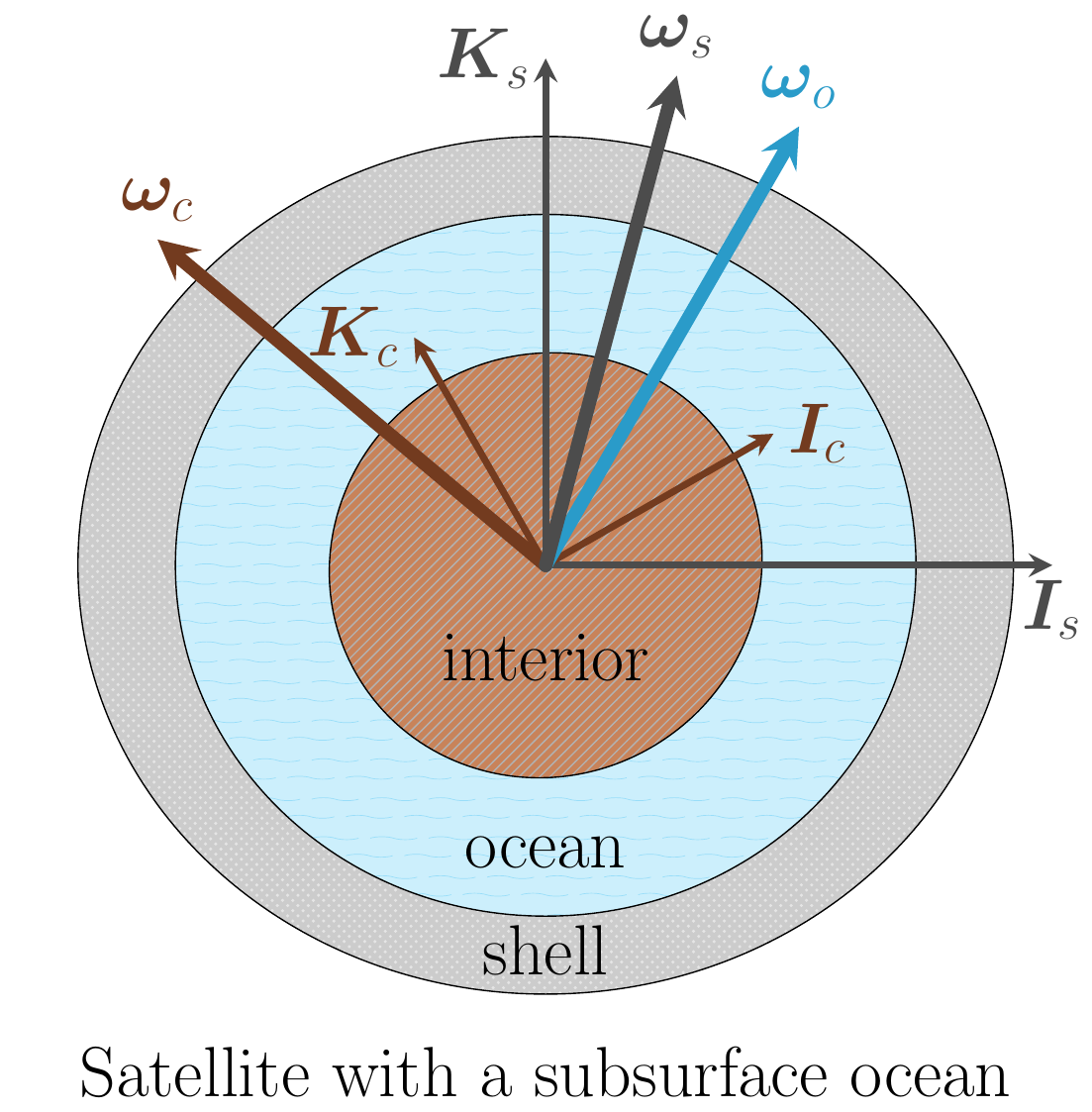}
\caption{\label{fig.models}Rigid satellites are characterised by their
basis vectors $(\vec I, \vec J, \vec K)$ and their rotation vector $\vec
\omega$ with respect to the laboratory frame. The same vectors are used
for satellites with a liquid core, but the angular speed $\vec
\omega'_c$ of the core with respect to the mantle is also specified. In
the case of a satellite with a global ocean, all vectors are expressed
in the laboratory frame. These are the basis vectors of the shell $(\vec
I_s ,\vec J_s, \vec K_s)$ and of the interior $(\vec I_c, \vec J_c,
\vec K_c)$, and the rotation vectors $\vec \omega_c$, $\vec \omega_o$,
$\vec \omega_s$ associated with the central region, the ocean and the
shell, respectively.}
\end{center}
\end{figure}
}

\newcommand{\figeigenmodes}{%
\begin{figure}
\begin{center}
\includegraphics[width=\linewidth]{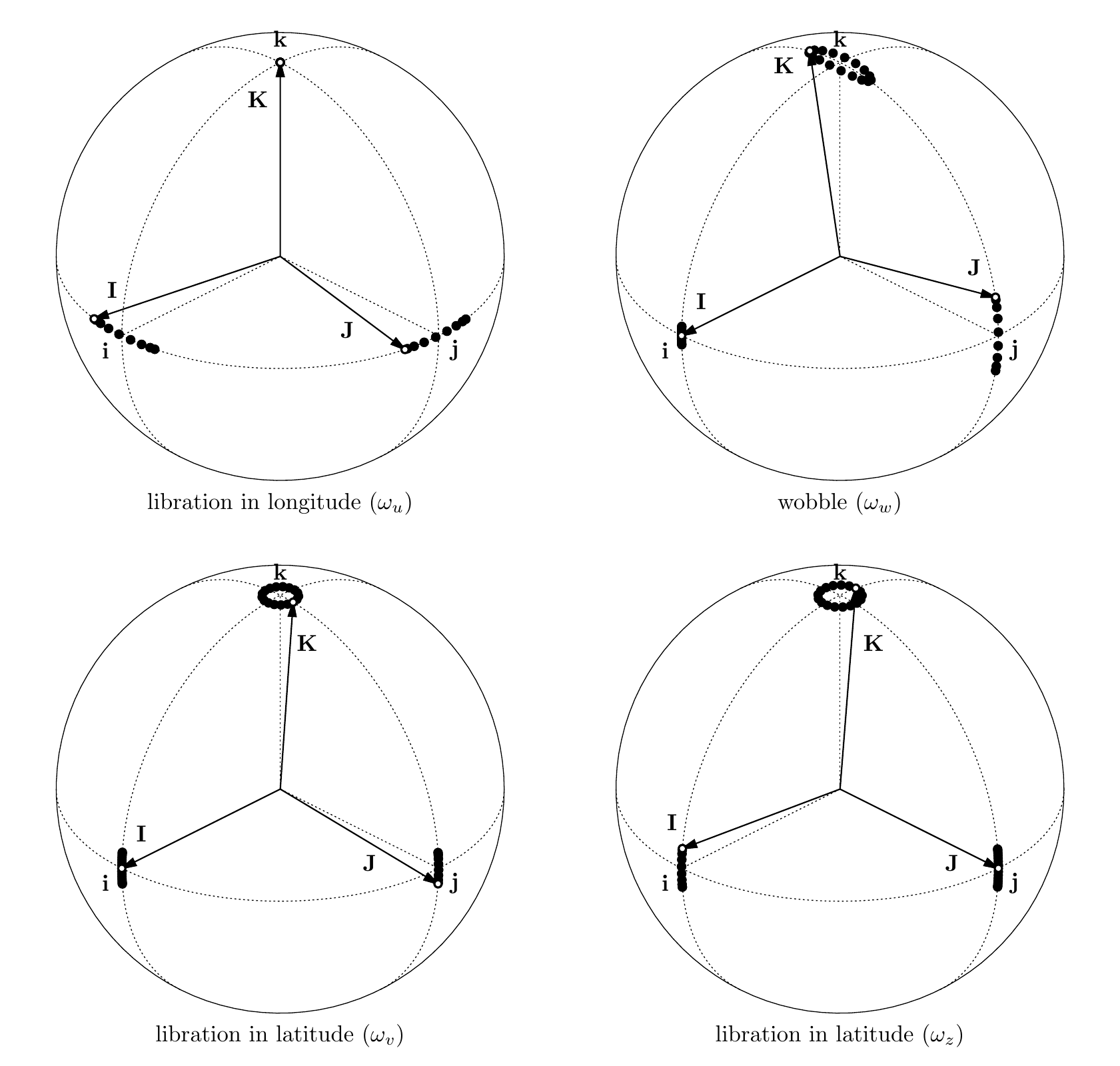}
\caption{\label{fig.em}Eigenmodes of Io's rotation motion computed with
the parameters of Tab.~\ref{tab.orbIo}. Positions at constant
time intervals of the principal axes $(\vec I, \vec J, \vec K)$ are
depicted by black dots. Open circles indicate the initial condition.
Intersections of the dotted great circles of the unit sphere represent
the laboratory frame $(\vec i, \vec j, \vec k)$.  Jupiter is in the
direction of the vector $\vec i$. The associated eigenfrequencies are
recalled below each figure.
}
\end{center}
\end{figure}
}
\newcommand{\figcistrans}{%
\begin{figure}
\includegraphics[width=\linewidth]{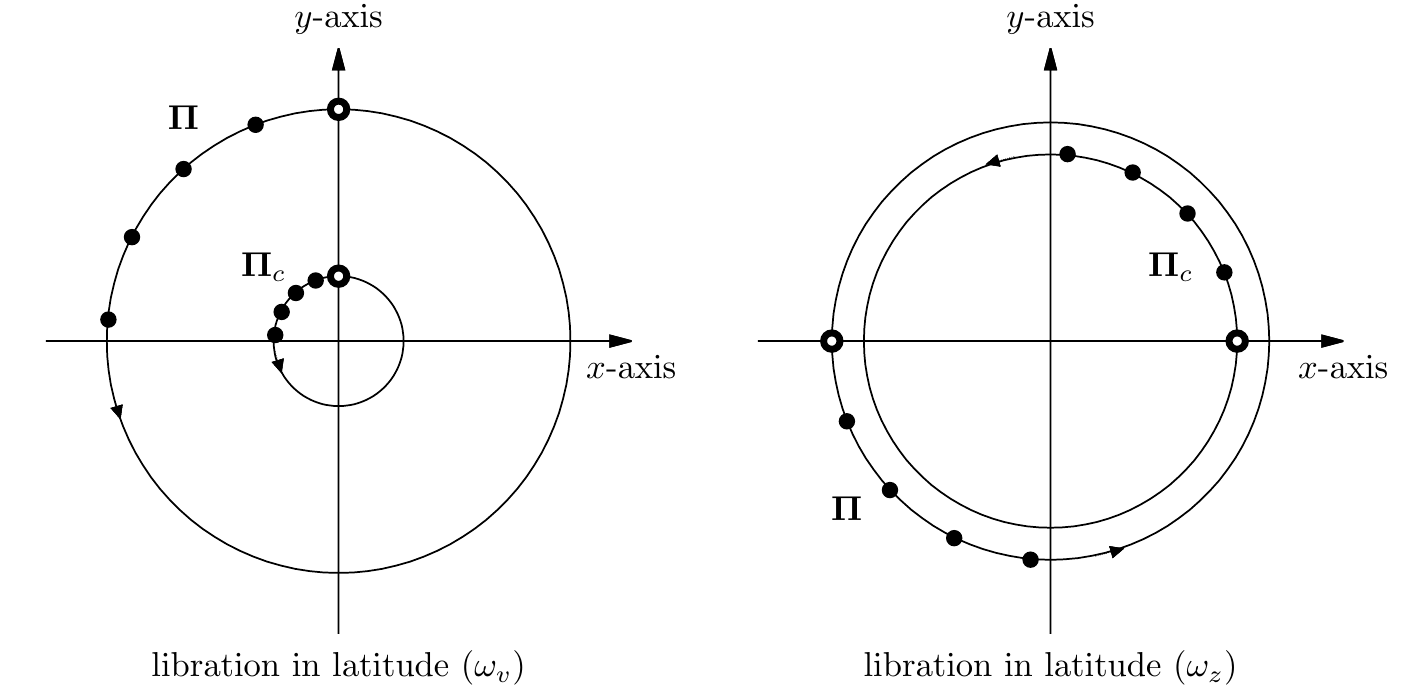}
\caption{\label{fig.cistrans}Trajectories of the projections of $\vec
\Pi$ and $\vec \Pi_c$ on the plane $(\vec i, \vec j)$ while Io is in
libration in latitude. Dots represent successive positions of the
vectors. Open circles denote the initial conditions. In the
eigenmode with frequency $\omega_v$, the two vectors are on the same
side from the origin whereas in the eigenstate of frequency $\omega_z$,
they are on opposite side. The radial coordinate of each vector is
plotted in a log scale with arbitrary units.  These figures have been
computed using Io's parameters (cf Tab.~\ref{tab.orbIo}).
}
\end{figure}
}
\newcommand{\figlibtitan}{%
\begin{figure}
\begin{center}
\includegraphics[width=\linewidth]{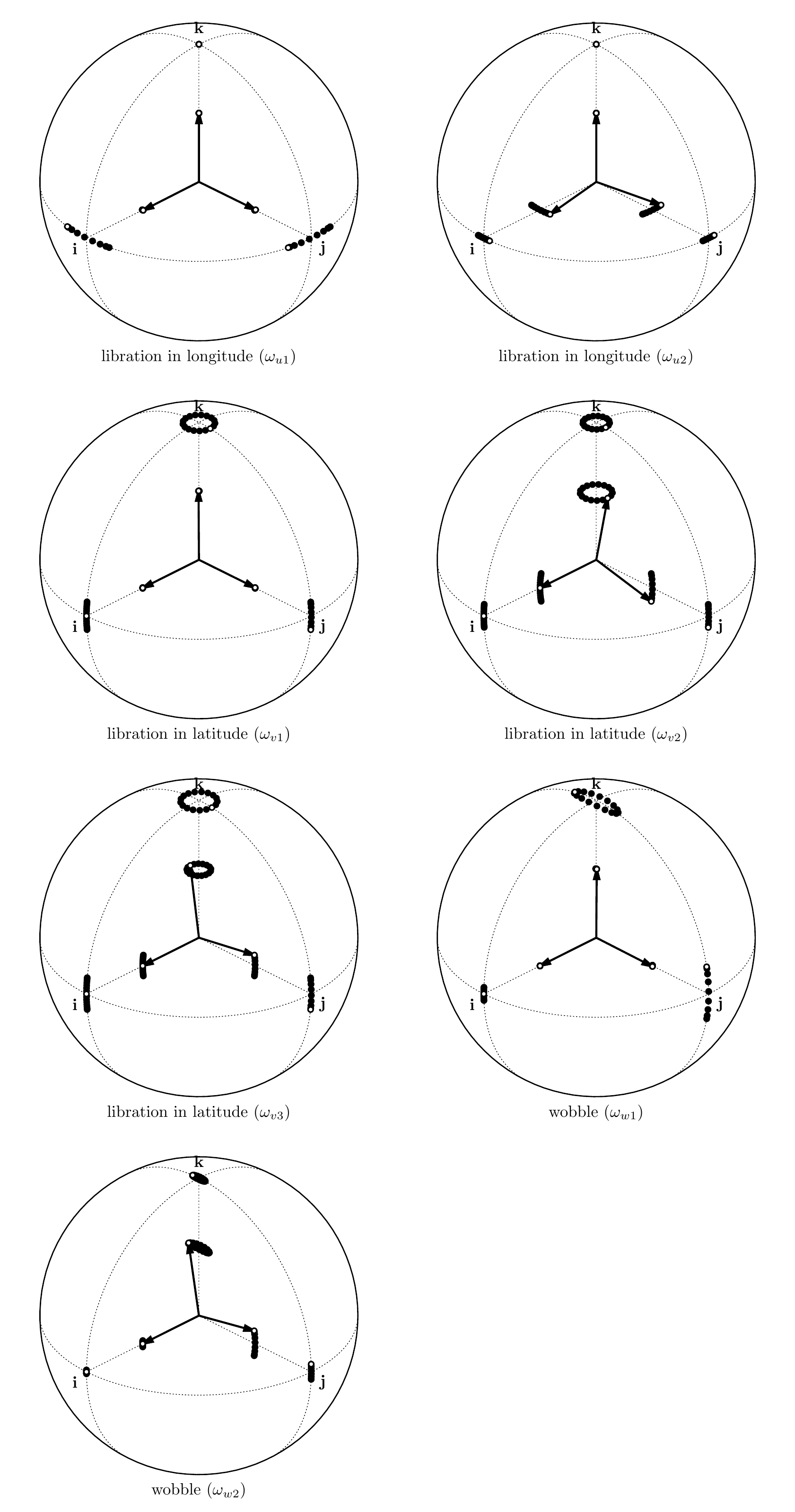}
\caption{\label{fig.libTitan}Eigenmodes of Titan's rotation motion
computed with the interior model F1. Positions at constant time
intervals of the shell principal axes $(\vec I_s, \vec J_s, \vec K_s)$
are depicted by black dots on the unit sphere. Those of the interior
$(\vec I_c, \vec J_c, \vec K_c)$ are plotted at half the radius of the
unit sphere. The white dots indicate the initial condition.
Intersections of the dotted great circles of the unit sphere represent
the laboratory frame $(\vec i, \vec j, \vec k)$.  Saturn is in the
direction of the vector $\vec i$. The associated eigenfrequencies are
recalled below each figure.
}
\end{center}
\end{figure}
}
\newcommand{\figpositiveobliq}{%
\begin{figure}
\begin{center}
\includegraphics[width=0.5\linewidth]{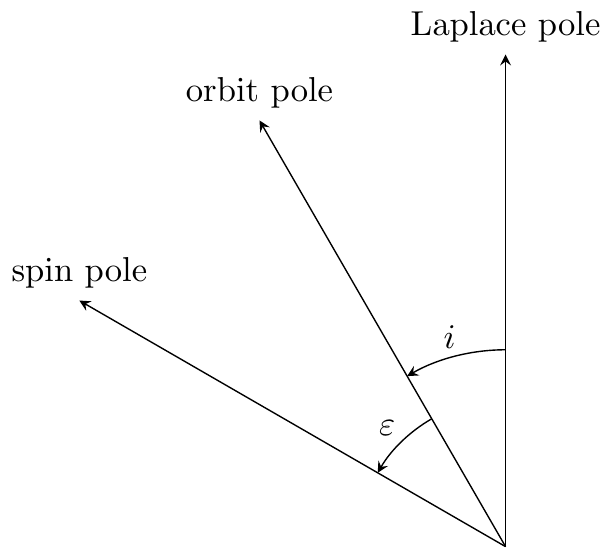}
\caption{\label{fig.po}Definition of Titan's inclination $i$ and obliquity
$\varepsilon$. In a Cassini state of the averaged problem, the Laplace
pole, the orbit pole and the spin pole are in a same plane. We define
the orientation of this plane by the inclination measured from the
Laplace pole to the orbit pole which by convention is positive. This
allows to defined the obliquity as a signed angle measured from the
orbit pole to the spin axis. In this figure, $\varepsilon$ is positive.}
\end{center}
\end{figure}
}
\newcommand{\figevolobliq}{%
\begin{figure}
\begin{center}
\includegraphics[width=\linewidth]{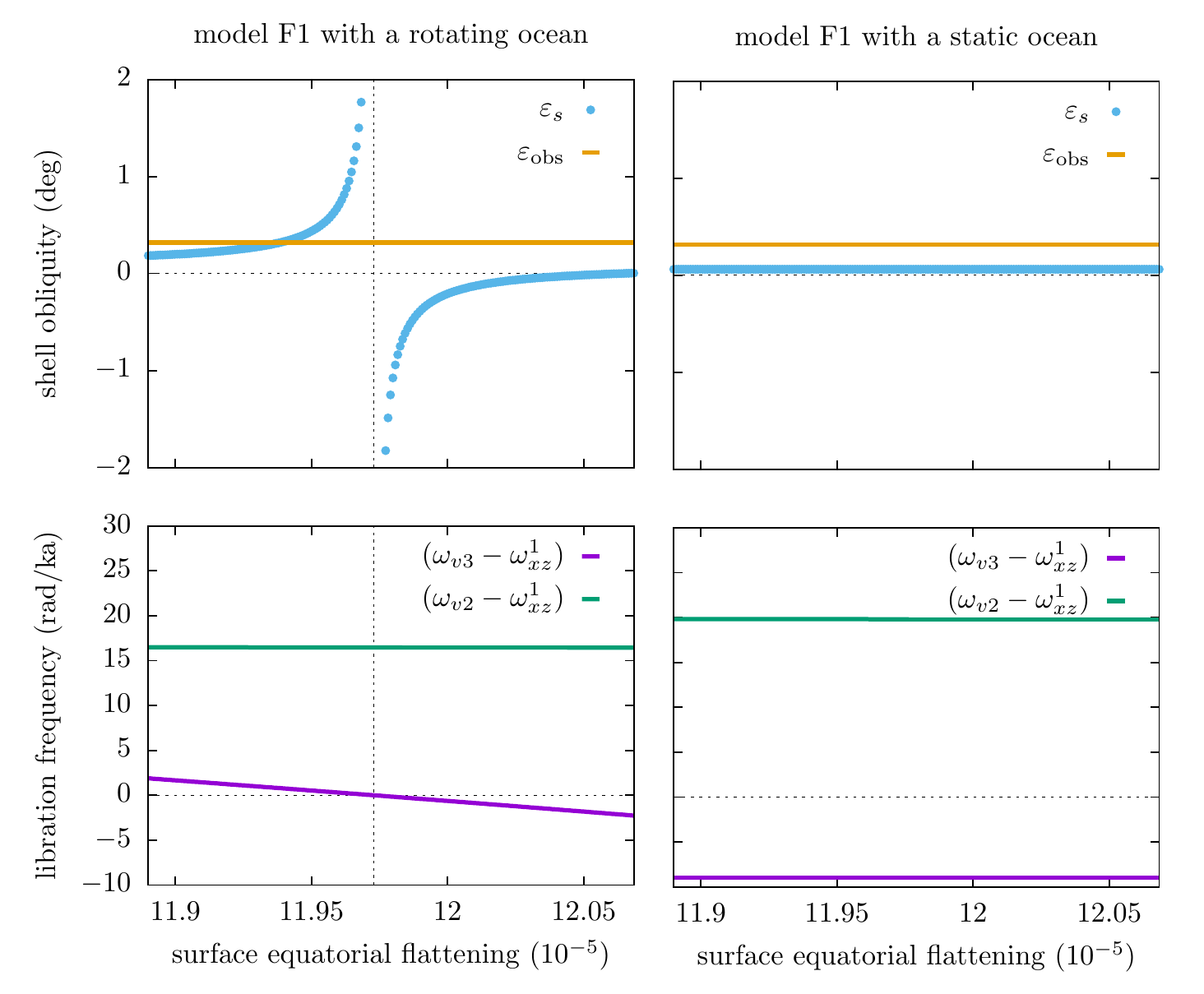}
\caption{\label{fig.obliq}Obliquity and libration frequencies as a
function of the surface equatorial flattening assuming a rotating ocean
(left) or a static ocean (right). The vertical dashed line indicates the
location of the resonance $\omega_{v3}=\omega^1_{xz}$. In the upper
plots, $\varepsilon_s$ represents the obliquity of the shell at the
Cassini state and $\varepsilon_\mathrm{obs}$ the observed value.
}
\end{center}
\end{figure}
}
\newcommand{\tabnotation}{%
\begin{table}
\begin{center}
\caption{\label{tab.notation}Notations}
\renewcommand{\arraystretch}{1.1}
\begin{tabular}{ll} \hline \hline
symbol & definition \\ \hline
${}_c,{}_o,{}_m,{}_s$ & indices standing for Core, Ocean, Mantle, and Shell, respectively \\
${}_\mathrm{rs},{}_\mathrm{fc},{}_\mathrm{go}$ & indices standing for
Rigid Satellite, Fluid Core, and Global Ocean \\ \hline
$\mathcal{F}_\mathrm{in} = (\vec i_0, \vec j_0, \vec k_0)$ & inertial frame \\
$\mathcal{F}_\mathrm{lab} = (\vec i, \vec j, \vec k)$ & laboratory frame \\
$\mathcal{F}_i = (\vec I_i, \vec J_i, \vec K_i)$ & frame associated with the layer $i$ \\
$\vec \Omega$ & rotation vector of $\mathcal{F}_\mathrm{lab}$ with respect to $\mathcal{F}_\mathrm{in}$ expressed in $\mathcal{F}_\mathrm{lab}$ \\ \hline
$\vec \omega_i$ & rotation vector of $\mathcal{F}_i$ with respect to $\mathcal{F}_\mathrm{lab}$ expressed in $\mathcal{F}_\mathrm{lab}$ \\
$\vec \omega'_c$ & rotation vector of $\mathcal{F}_c$ with respect to $\mathcal{F}_m$ expressed in $\mathcal{F}_m$ \\
$\vec \Pi_i$ & Lie momentum associated with $\vec \omega_i$ \\
$\vec \Pi'_c$ & Lie momentum associated with $\vec \omega'_c$ \\
$\vec I_i, \vec J_i, \vec K_i$ & basis vectors of $\mathcal{F}_i$ expressed in $\mathcal{F}_\mathrm{lab}$ \\
$\mat R_i = [\vec I_i, \vec J_i, \vec K_i]$ & rotation matrix of the layer $i$ relative to $\mathcal{F}_\mathrm{lab}$ \\
$\vec y_i = (\vec \Pi_i, \vec I_i, \vec J_i, \vec K_i)$ & state vector of the layer $i$ \\
$\vec y$ & state vector of the whole system \\ \hline
$T(\vec y)$ & kinetic energy \\
$U(\vec y, t)$ & potential energy \\
$L(\vec y,t)$ & Lagrangian \\
$H(\vec y, t)$ & Hamiltonian \\
$C_i(\vec y)$ & Casimir functions \\
$\mu_i$ & Lagrange multipliers \\
$F(\vec y)$ & Lagrangian associated with the minimisation of $H_0$ with constraints \\
$\mat B(\vec y)$ & Poisson matrix \\
$\mat A(\vec y)$ & matrix of the linearised system \\ \hline
$U_0(\vec y)$ & constant part of $U(\vec y, t)$ \\
$U_1(\vec y, t)$ & perturbation $U(\vec y,t) - U_0(\vec y)$ \\
$U_\mathrm{self}(\vec y)$ & self gravitational energy of the satellite \\
$(u_{ij})_{i,j\in\{x,y,z\}}$ & constant parameters of $U_\mathrm{self}$ \\
$H_0(\vec y)$ & autonomous part of $H(\vec y, t)$ \\
$H_1(\vec y,t)$ & perturbation $H(\vec y, t) - H_0(\vec y)$ \\ \hline
$\vec r$, $\vec r(t)$ & radius vector connecting the satellite barycenter to the planet \\
$\mat S(t)$ & ${\cal G}M_p\vec r \trans{\vec r}/r^5$ \\
$\mat S_0$ & constant part of $\mat S(t)$ \\
$\mat S_1(t)$ & $\mat S(t) - \mat S_0$ \\
$(\sigma^0_{uv})_{u,v\in\{x,y,z\}}$ & elements of the matrix $\mat S_0$ \\
$(\sigma^1_{uv}(t))_{u,v\in\{x,y,z\}}$ & elements of the matrix $\mat S_1(t)$ \\ \hline
${\cal G}$ & gravitational constant \\
$M_p$ & mass of the central planet \\
$\alpha_i$, $\beta_i$, $\gamma_i$ & $(C_i-B_i)/A_i$, $(C_i-A_i)/B_i$, $(B_i-A_i)/C_i$, respectively \\
$\rho_i$ & density of the layer $i$ \\
$a_i, b_i, c_i$ & radii of the outer boundary of the layer $i$ \\
$\zeta$ & equatorial flattening $(a-b)/a$  \\
$\inertia_i$ & inertia tensor of the layer $i$ expressed in
$\mathcal{F}_\mathrm{lab}$ \\
$\inertia'$ & ancillary inertia tensor \\
$A_i, B_i, C_i$ & principal moments of inertia of the layer $i$ \\
$A', B', C'$ & ancillary moments of inertia \\
$\omega_u$ & frequency of libration in longitude \\
$\omega_v$ & frequency of libration in latitude \\
$\omega_w$ & wobble frequency \\
\hline
\end{tabular}
\end{center}
\end{table}
}
\newcommand{\taborbio}{%
\begin{table}
\begin{center}
\caption{\label{tab.orbIo} Orbital and physical parameters of Io taken from
\citep{Noyelles14}.}
\begin{tabular}{lrl} \hline\hline
Parameter & value & units \\ \hline
${\cal G}M_p$ (Jupiter) & 126\,712\,765 & km$^3$/s$^{2}$ \\
$a$            & 422\,029.958 & km \\
$e$            & 0.00415 & \\
$i$            & 2.16 & arcmin \\
$\Omega$       & 1297.204\,472\,527\,9755 & rad/\Year \\
$A/(mR^2)$     & 0.375\,127 & \\
$B/(mR^2)$     & 0.377\,342 & \\
$C/(mR^2)$     & 0.378\,080 & \\
$A_c/(mR^2)^a$   & 0.006\,007\,5578 & \\
$B_c/(mR^2)^a$   & 0.006\,283\,9600 & \\
$C_c/(mR^2)^a$   & 0.006\,253\,4432 & \\
\hline
\end{tabular} \\
{\scriptsize
$^a$ Moments of inertia of the core computed from the internal model 1 of
\citep{Noyelles14}.}
\end{center}
\end{table}
}
\newcommand{\tabfreqio}{%
\begin{table}
\begin{center}
\caption{\label{tab.freqIo} Eigenperiods of Io's rotational motion
(Eq.~\ref{eq.Tuvwz}).}
\begin{tabular}{lrrrr} \hline\hline
source & $T_u$ (day) & $T_v$ (day) & $T_w$ (day) & $T_z$ (day) \\ \hline
\citet{Noyelles14}
& 13.2322     & 166.3520    & 225.0927    & 1.7382      \\ 
This work: model fc\ \  (Sect.~\ref{sec.poincare-hough})
& 13.2504     & 157.2780    & 224.5395    & 1.7385      \\
This work: model fc$^\prime$ (Sect.~\ref{sec.fluid-approximation})
& 13.2502     & 156.5653    & 224.5402    & 1.7368      \\
\hline
\end{tabular}
\end{center}
\end{table}
}

\newcommand{\taborbtitan}{%
\begin{table}
\begin{center}
\caption{\label{tab.orbTitan} Orbital parameters of Titan used in this
study.}
\begin{tabular}{lrll} \hline\hline
Parameter & value & units & reference \\ \hline
${\cal G}M_p$ (Saturn)  & $37\,931\,272$ & km$^3$/s$^{2}$ & \citep{Campbell89}\\
$a$            & 1\,221\,729 & km & computed$^a$\\
$e$            & 0.028 & & \citep{Vienne95} \\
$i^b$          & 0.320 & deg & \citep{Vienne95} \\
$\Omega$       & 143.924\,047\,85 & rad/\Year & \citep{Vienne95} \\
$\dd\Phi/\dd t$     & -0.008\,931\,24  & rad/\Year & \citep{Vienne95} \\
\hline
\end{tabular}
\end{center}
{\scriptsize
$^a$ The semimajor axis has been computed from the masses of Saturn and
Titan given by \citet{Campbell89} and the orbital parameters $N_6$ and
$p_{06}$ provided by \citet{Vienne95}. \\
$^b$ Inclination with respect to the Laplace plane given by the
amplitude of the second harmonic of $\zeta_{06}$ in the notation of
\citet{Vienne95}.
}
\end{table}
}
\newcommand{\tabphystitan}{%
\begin{table}
\begin{center}
\caption{\label{tab.physTitan}Physical parameters of the two
interior models of Titan considered in this study taken from
\citep{Fortes12}.}
\begin{tabular}{l|*3{c}|*3{c}} \hline\hline
\multicolumn{1}{c}{}
 & \multicolumn{3}{c}{\bf F1} 
 & \multicolumn{3}{c}{\bf F2}  \\ \hline
      & $\rho$            & $R$      & $\zeta$ 
      & $\rho$            & $R$      & $\zeta$ \\ 
Layer & \footnotesize (kg/m$^3)$ & \footnotesize (km) & \footnotesize $(10^{\text{-}5})$  
      & \footnotesize (kg/m$^3)$ & \footnotesize (km) & \footnotesize $(10^{\text{-}5})$ \\ \hline
Ice & 930.9 & 2575 & 12.068
    & 930.9 & 2575 & 12.080 \\
Ocean & 1023.5 & 2475 & 11.878
      & 1281.3 & 2475 & 11.887 \\
Ice V  & 1272.7 & 2225 & 11.552
       & 1350.9 & 2225 & 11.488 \\
Ice VI & 1338.9 & 2163 & 11.521
       & - & - & - \\
Silicate & 2542.3 & 2116 & 11.514
         & 2650.4 & 1984 & 11.310 \\ \hline
\end{tabular}
\end{center}
\scriptsize
For each layer, $\rho$ is the density and $R$ and $\zeta$ respectively
denote the mean radius and the equatorial flattening of the upper
boundary.
\end{table}
}
\newcommand{\tabderivtitan}{%
\begin{table}
\begin{center}
\caption{\label{tab.derivTitan}Derived parameters for Titan's model.}
\renewcommand{\arraystretch}{1.3}
\begin{tabular}{lrrl} \hline\hline
Parameter & model {\bf F1} & model {\bf F2}  & units\\ \hline
$A_c/(mR^2)$    & 0.232\,133\,9588 & 0.213\,354\,6838 &           \\
$B_c/(mR^2)$    & 0.232\,160\,7420 & 0.213\,379\,0654 &           \\
$C_c/(mR^2)$    & 0.232\,169\,6677 & 0.213\,387\,1908 &           \\ \hline
$A_m/(mR^2)$    & 0.035\,565\,0464 & 0.035\,556\,8942 &           \\
$B_m/(mR^2)$    & 0.035\,569\,6492 & 0.035\,561\,5041 &           \\
$C_m/(mR^2)$    & 0.035\,571\,1830 & 0.035\,563\,0404 &           \\ \hline
$A'_c/(mR^2)$   & 0.104\,835\,1592 & 0.131\,211\,1289 &           \\
$B'_c/(mR^2)$   & 0.104\,847\,2721 & 0.131\,226\,2055 &           \\
$C'_c/(mR^2)$   & 0.104\,851\,3089 & 0.131\,231\,2299 &           \\ \hline
$A'_m/(mR^2)$   & 0.178\,538\,4650 & 0.223\,457\,6674 &           \\
$B'_m/(mR^2)$   & 0.178\,559\,6760 & 0.223\,484\,2365 &           \\
$C'_m/(mR^2)$   & 0.178\,566\,7448 & 0.223\,493\,0909 &           \\ \hline
$u_{xx}/(mR^2)$ & 135.969\,642\,03 & 109.837\,900\,34 & 1/day$^{2}$\\
$u_{xy}/(mR^2)$ & 135.989\,307\,93 & 109.853\,755\,74 & 1/day$^{2}$\\
$u_{xz}/(mR^2)$ & 135.995\,863\,22 & 109.859\,040\,86 & 1/day$^{2}$\\
$u_{yx}/(mR^2)$ & 135.938\,311\,45 & 109.813\,022\,76 & 1/day$^{2}$\\
$u_{yy}/(mR^2)$ & 135.957\,972\,82 & 109.828\,874\,57 & 1/day$^{2}$\\
$u_{yz}/(mR^2)$ & 135.964\,526\,60 & 109.834\,158\,49 & 1/day$^{2}$\\
$u_{zx}/(mR^2)$ & 135.927\,870\,23 & 109.804\,732\,03 & 1/day$^{2}$\\
$u_{zy}/(mR^2)$ & 135.947\,530\,09 & 109.820\,582\,64 & 1/day$^{2}$\\
$u_{zz}/(mR^2)$ & 135.954\,083\,36 & 109.825\,866\,17 & 1/day$^{2}$\\
\hline
\end{tabular}
\end{center}
\scriptsize
Note that the number of digits provided in this table is required to recover the
values presented in Tabs.~\ref{tab.eigenfreq} and \ref{tab.obliq}.
\end{table}
}
\newcommand{\tabeigenfreq}{%
\begin{table}
\begin{center}
\caption{\label{tab.eigenfreq}Eigenfrequencies of Titan's rotation in
rad/\Year.}
\renewcommand{\arraystretch}{1.2}
\begin{tabular}{l|rr|rr|r|p{1.9cm}} \hline \hline
 & \multicolumn{2}{c|}{rotating ocean} & \multicolumn{2}{c|}{static ocean} & \multicolumn{1}{c|}{rigid} \\
 & {\bf F1} & {\bf F2} & {\bf F1} & {\bf F2} & {\bf F1}/{\bf F2} & type of motion \\ \hline
$\omega_{u1}$ &   7.9237 &   8.2656 &   7.9237 &   8.2656 & \multirow{2}{*}{2.7117} & libration in \\
$\omega_{u2}$ &   2.3950 &   2.1147 &   2.3950 &   2.1147 &                         & longitude \\
\hline
$\omega_{v1}$ & 144.3272 & 144.3641 & 144.2507 & 144.2683 &          & \multirow{3}{1.9cm}{libration in latitude} \\
$\omega_{v2}$ & 143.9494 & 143.9445 & 143.9528 & 143.9472 & 143.9582 & \\
$\omega_{v3}$ & 143.9307 & 143.9266 & 143.924$^a$ & 143.924$^a$ &          & \\
\hline
$\omega_{w1}$ &   0.1943 &   0.2105 &   0.1177 &   0.1104 & \multirow{2}{*}{0.0228} & \multirow{2}{*}{wobble} \\
$\omega_{w2}$ &   0.0178 &   0.0138 &   0.0214 &   0.0199 &                         & \\
\hline
\end{tabular}
\end{center}
$^a$ In the case where the ocean is assumed static,
$\omega_{v3}=143.9240$ rad/a is the mean motion $\Omega$.
\end{table}
}
\newcommand{\tabobliq}{%
\begin{table}
\begin{center}
\caption{\label{tab.obliq}Obliquity of Titan's layers in degree.}
\renewcommand{\arraystretch}{1.2}
\begin{tabular}{l|rr|rr|r} \hline \hline
 & \multicolumn{2}{c|}{rotating ocean} & \multicolumn{2}{c|}{static ocean} & \multicolumn{1}{c}{rigid} \\
 & {\bf F1} & {\bf F2} & {\bf F1} & {\bf F2} & {\bf F1}/{\bf F2}  \\ \hline
core  &   0.294  &  0.272 &   0.149  &   0.207  & 0.113 \\
ocean &  -0.479  &  0.208 &  -0.320  &  -0.320  & 0.113 \\
shell &   0.004  &  0.108 &   0.062  &   0.064  & 0.113 \\
\hline
\end{tabular}
\end{center}
{\scriptsize 
The meaning of the sign of the obliquity is explained in
Fig.~\ref{fig.po}.
}
\end{table}
}

\begin{document}

\title{Rotation of a rigid satellite with a fluid component}
\subtitle{A new light onto Titan's obliquity}

\author{Gwena\"el BOU\'E$^1$ \and Nicolas RAMBAUX$^1$ \and Andy RICHARD$^{1,2}$}

\authorrunning{G. Bou\'e, N. Rambaux \& A. Richard} 

\institute{Gwena\"el Bou\'e  \at
              \email{Gwenael.Boue@obspm.fr}
           \and
           Nicolas Rambaux \at
              \email{Nicolas.Rambaux@obspm.fr}
           \and
           Andy Richard \at
              \email{Andy.Richard@universcience.fr}
           \and
           $^1$ IMCCE, Observatoire de Paris, UPMC Univ. Paris 6, PSL Research University, Paris, France \\
           $^2$ Universcience, Palais de la d\'ecouverte, Paris, France
}

\date{Received: date / Accepted: date \\ \large Version: \today}

\graphicspath{ {figures/} }

\maketitle

\begin{abstract}
We revisit the rotation dynamics of a rigid satellite with either a liquid core or a global
sub-surface ocean. In both problems, the flow of the fluid component is assumed inviscid. The study
of a hollow satellite with a liquid core is based on the Poincar\'e-Hough model which provides exact
equations of motion. We introduce an approximation when the ellipticity of the cavity is low. This
simplification allows to model both types of satellite in the same manner. The analysis of their
rotation is done in a non-canonical Hamiltonian formalism closely related to Poincar\'e's ``forme
nouvelle des \'equations de la m\'ecanique''. In the case of a satellite with a global ocean, we
obtain a seven-degree of freedom system.  Six of them account for the motion of the two rigid
components, and the last one is associated with the fluid layer. 
We apply our model to Titan for which the origin of the obliquity is still a debated question. We
show that the observed value is compatible with Titan slightly departing from the hydrostatic
equilibrium and being in a Cassini equilibrium state.

\end{abstract}


\section{Introduction}
\label{sec.intro}
The spin pole of Titan, Saturn's largest moon, is lying close to the plane defined by its orbit pole
and the Laplace pole \citep{Stiles08,Stiles10}. This observation, made by the RADAR instrument of
the Cassini mission, suggests that Titan is in (or very close to) a Cassini state \citep{Colombo66,
Peale69}. For a rigid body, the equilibrium obliquity is a function of its moments of inertia. Those
of Titan have been deduced from its Stokes coefficients $J_2= (33.599\pm 0.332)\times10^{-6}$ and
$C_{22}=(10.121 \pm 0.029)\times10^{-6}$ and from the hydrostatic equilibrium hypothesis implying a
mean moment of inertia $I/(mR^2) = 0.3431$ \citep[SOL1a]{Iess12}, where $m$ and $R$ are the mass and
radius of Titan, respectively. The assumed hydrostatic equilibrium is suggested by
the ratio $J_{2}/C_{22}\approx 10/3$ which is precisely the expected value for a hydrostatic body
\citep[e.g.,][]{Rappaport97}. Assuming these values, if Titan were rigid and in a Cassini
equilibrium state, its obliquity would be 0.113 deg \citep{Bills11}, i.e. about one third of the
radiometric value 0.32 deg \citep{Stiles08, Stiles10, Meriggiola16}. To match the
observations, the frequency of the free libration in latitude must be reduced by a factor 0.526
\citep{Bills11}. In particular, this would be the case if $I/(mR^2)$ were increased to 0.45 ({\em
ibid.}), a value exceeding 2/5 obtained for a homogeneous body, as if the mass of the satellite was
concentrated toward the surface. This result leads to think that the observed obliquity is that of a
thin shell partially decoupled from the interior by, e.g., a global ocean ({\em ibid.}).

The idea that the ice-covered satellites of the outer planets hold a global underneath ocean has
already been proposed based on models of their internal structures \citep[e.g.,][]{Lewis71}. Even
the dwarf planet Pluto is suspected to harbour a subsurface ocean \citep{Nimmo16}.  In the case of
Titan, the presence of the ocean is also revealed by laboratory experiments on the behaviour of
water-ammonia compounds at high pressure and low temperature \citep{Grasset96},
by the detection of electromagnetic waves in its atmosphere \citep{Beghin12} and by the high value
of its Love number $k_2$ \citep{Iess12}.

A dynamical problem closely related to the present one is that of a hollow satellite with a liquid
core as described by the Poincar\'e-Hough model \citep{Poincare10,Hough95}. For this specific
problem, \citet{Poincare01} developed a new Lagrangian formalism, based on the properties of the Lie
group acting on the configuration space, which allows to derive the equations of motion in a very
simple and elegant manner. Such a system is characterised by four degrees of freedom, three of them
being associated with the rotation of the rigid mantle and the last one being due to the
motion of the liquid core \citep[e.g.,][]{Henrard08}. Applying this model to Jupiter's satellite Io,
\citet{Henrard08} observed that the frequency of the additional degree of freedom is close to the
orbital frequency and should thus multiply the possibility of resonances. For Titan, we shall expect
the same conclusion due to the presence of the ocean, but unfortunately, \citeauthor{Poincare10}'s
model relies on the concept of a fluid simple motion which cannot be rigorously transposed to the
case of a satellite with a global subsurface ocean.

In the case of Titan, the effect of an ocean on the rotation dynamics has been studied numerically
using Euler's rotation equations taking into account the gravitational interaction of Saturn on each
layer, the pressure torques at the two fluid-solid boundaries, and the gravitational coupling
between the interior and the shell \citep{Baland11, Baland14, Noyelles14b}. The elastic deformation
of the solid layers and the atmospheric pressure have also been included in a modelling of the
libration in longitude \citep{Richard14b} and in a modelling of the Chandler polar motion
\citep{Coyette16}.  Despite several arguments in favour of an ocean, this model does not easily
explain the tilt of Titan's spin-axis. Indeed, under the hydrostatic equilibrium hypothesis,
\citet{Baland11} and \citet{Noyelles14b} found that the obliquity of the Cassini state remains
bounded below 0.15 deg, i.e. about one half of the observed value. There thus seemed to be a need
for a significant resonant amplification to bring the system out of the Cassini equilibrium
\citep{Baland11,Noyelles14b}.  However these studies do not invoke the same mode as the origin of
the resonant amplification.  In addition, this solution does not agree with extended observations of
the spin-axis orientation \citep{Meriggiola12}.  The model has then been amended to allow the
Cassini state obliquity to reach the observed 0.32 deg, but this has only been made possible after
releasing the hydrostatic shape assumption leaving the ratio $J_2/C_{22} \approx 10/3$ unexplained
\citep{Baland14}.

It should be stressed that models developed thus far discard the rotation of the ocean relative to
the inertial frame. This is a valid assumption to reproduce librations in longitude
\citep[e.g.,][]{Richard14}, but not anymore for precession motion. By consequence, the associated
dynamical system only has 6 degrees of freedom equally shared by the rigid interior and the shell
\citep{Noyelles14b}. Yet, a comparison of this problem with that of a satellite with a liquid core
strongly suggests that a three layered body must have 7 degrees of freedom, one of which being
brought by the ocean. Here, we aim at building a new dynamical model accounting for the rotation of
the liquid layer as done by \citet{Mathews91} for the Earth. More recently, the latter model
has been adapted to the study of the Moon \citep{Dumberry16}\footnote{ \citet{Dumberry16} could only
highlight 5 degrees of freedom because their model of the Moon is axisymmetric and not triaxial.}
and of Mercury \citep{Peale16}. Here we reconsider the problem with a Hamiltonian approach. In that
scope, we first extend the Lagrangian formalism described in \citep{Poincare01} to a non-canonical
Hamiltonian formalism allowing to study relative equilibria in a very efficient manner as in
\citep{Maddocks91,Beck98}.  The method has proven its efficiency in the context of a rigid satellite
in circular orbit \citep{Beck98}, in the analysis of the two rigid body problem
\citep{Maciejewski95}, and in several studies of the attitude of a satellite with a gyrostat
\citep[e.g.,][and references therein]{Hall07,Wang12}. The approach is described in
Sect.~\ref{sec.non-canonical} and illustrated in the case of a rigid satellite in
Sect.~\ref{sec.rigid-satellite}. We revisit the problem of a moon with a fluid core with this
approach and we propose a simplification straightforwardly transposable to a three layered body in
Sect.~\ref{sec.liquid-core}. The rotation dynamics of a satellite with a subsurface ocean is
presented in Sect.~\ref{sec.ocean}. In the subsequent section~\ref{sec.application}, we test our
model and our simplification on Io, a satellite with a liquid core, verifying that the derived
eigenfrequencies are in very good agreement with those obtained in previous studies of the same
problem made by \citet{Noyelles13,Noyelles14}. In this section, we also analyse the case of Titan
showing that the additional degree of freedom makes the system highly sensitive to the internal
structure and that the observed obliquity can be easily reproduced. Finally, we discuss our model
and conclude in Sect.~\ref{sec.conclusion}. The notation used in this paper is explained in
Tab.~\ref{tab.notation}.

\section{Non-canonical Hamiltonian formalism}
\label{sec.non-canonical}

\subsection{Equations of motion}
\label{sec.poincare-hamilton}
\subsubsection{General case}
Let a dynamical system with $n$ degrees of freedom described by a Lagrangian $L$. We denote by $Q$
the configuration space and each point $\vec q\in Q$ is represented by a set of $m\geq n$
coordinates $(q_1,\cdots,q_m)$. The number of coordinates is purposely allowed to be greater than
the actual dimension of the manifold $Q$.  As in \citep{Poincare01}, we assume that there exists a
transitive Lie group $G$ acting on $Q$. The transitivity of $G$ means that for all $\vec q, \vec
q'\in Q$, there exists an element $g$ of the group $G$ such that $\vec q' = g\vec q$. In particular,
given an initial condition $\vec q_0$, there exists $g_t \in G$ such that the configuration $\vec
q(t)$ at time $t$ reads $\vec q(t) = g_t \vec q_0$. In this work, $G$ will be the rotation group
$SO(3)$, the translation group $T(3)$, or some combinations of both.

Let $\mathfrak g$ be the Lie algebra of $G$. By definition, there exists $\field X\in \mathfrak g$
such that the generalised velocity reads $\dot{\vec q} = \field X(\vec q)$. Since the action of $G$
on $Q$ is transitive, the dimension of $\mathfrak g$ is equal to the number $n$ of degrees of
freedom. Let ${\mathcal{B}} = (\field X_1, \cdots, \field X_n)$ be a basis of $\mathfrak g$ and
$(X_{ij})_{1\leq i\leq n, 1\leq j\leq m}$ be the $n\times m$ functions of $\vec q$ defined as
\begin{equation}
\field X_i = \sum_{j=1}^m X_{ij} \frac{\partial }{\partial
q_j} .
\label{eq.Xij}
\end{equation}
We denote by $\vec \eta = (\eta_1, \cdots, \eta_n) \in \mathbb{R}^n$ the coordinates of $\field X$
in $\mathcal{B}$ such that
\begin{equation}
\dot{\vec q} = \sum_{i=1}^n \eta_i \field X_i(\vec q)\ .
\label{eq.dotq}
\end{equation}
Because the term ``generalised velocity'' is already attributed to $\dot{\vec q}$, hereafter we call
$\vec \eta$ the {\em Lie velocity} of the system. Given two configurations $\vec q$ and $\vec q'$
infinitely closed to each other, we also define the $n$-tuple $\delta \vec \xi = (\delta \xi_1,
\cdots, \delta \xi_n)$ such that
\begin{equation}
\delta \vec q \equiv \vec q' - \vec q = \sum_{i=1}^n \field X_i(\vec q)
\delta \xi_i .
\end{equation}
Poincar\'e considers the Lagrangian as a function of $(\vec \eta, \vec q)$ and writes its
infinitesimal variation as
\begin{equation}
\delta {L} = \sum_{i=1}^n \frac{\partial {L}}{\partial \eta_i} \delta \eta_i 
+ \field X_i({L}) \delta \xi_i .
\label{eq.dL}
\end{equation}
The resulting equations of motion are \citep{Poincare01}
\begin{equation}
\frac{\dd}{\dd t}\frac{\partial {L}}{\partial \eta_i} 
= \sum_{j,k} c_{ij}^k \eta_j \frac{\partial {L}}{\partial \eta_k}
+ \field X_i({L}),
\label{eq.Poincare}
\end{equation}
where $c_{ij}^k$, defined as
\begin{equation}
[\field X_i, \field X_j] \equiv
\field X_i \field X_j - \field X_j \field X_i = 
\sum_{k=1}^n c_{ij}^k \field X_k ,
\end{equation}
are the structure constants of $\mathfrak g$ with respect to the chosen basis $\mathcal B$.

To get the Hamiltonian equations equivalent to Eq.~(\ref{eq.Poincare}), we introduce a momentum
$\vec \pi$ associated with the Lie velocity $\vec \eta$, and defined as
\begin{equation}
\vec \pi \equiv \frac{\partial {L}}{\partial \vec \eta}.
\label{eq.defpi}
\end{equation}
Following the same nomenclature as for $\vec \eta$, we call this momentum $\vec \pi$ the {\em Lie
momentum} of the system.  The Hamiltonian ${H}$ is constructed by means of a Legendre transformation
as
\begin{equation}
{H}(\vec \pi, \vec q) \equiv \vec \pi \cdot \vec \eta - {L}(\vec
\eta, \vec q).
\label{eq.Hlegendre}
\end{equation}
Using Eqs.~(\ref{eq.dL}) and (\ref{eq.defpi}), the infinitesimal variation of ${H}$
(Eq.~\ref{eq.Hlegendre}) reads
\begin{equation}
\delta{H} = \sum_{i=1}^n \eta_i \delta \pi_i - \field X_i({L})\delta \xi_i.
\label{eq.dH1}
\end{equation}
But since ${H}$ is a function of $\vec \pi$ and $\vec q$, we also have, as in Eq.~(\ref{eq.dL}),
\begin{equation}
\delta {H} = \sum_{i=1}^n \frac{\partial {H}}{\partial \pi_i}
\delta \pi_i + \field X_i({H}) \delta \xi_i.
\label{eq.dH2}
\end{equation}
The identification of Eqs.~(\ref{eq.dH1}) and (\ref{eq.dH2}) gives
\begin{equation}
\eta_i = \frac{\partial {H}}{\partial \pi _i}
\qquad\mathrm{and}\qquad
\field X_i({H}) = -\field X_i({L}).
\end{equation}
Using these identifications, the expression of $\dot{\vec q}$ (Eq.~\ref{eq.dotq}), and Poincar\'e's
equation (\ref{eq.Poincare}) where $\partial {L}/\partial \eta_i$ is replaced by $\pi_i$
(Eq.~\ref{eq.defpi}), we get the non-canonical equations of motion associated with ${H}$,
viz.,
\begin{equation}
\dot q_i = \sum_{j=1}^n 
    \frac{\partial {H}}{\partial \pi_j} 
    \field X_j(q_i)
\qquad\mathrm{and}\qquad
\dot \pi_i = \sum_{j,k} c_{ij}^k \frac{\partial {H}}{\partial
\pi_j} \pi_k - \field X_i({H}).
\label{eq.Hamilton}
\end{equation}
Let us denote the state vector by $\vec y = (\vec \pi, \vec q) \in \mathbb{R}^{n+m}$. The equations
of motion (\ref{eq.Hamilton}) written in matrix form read
\begin{equation}
\dot {\vec y} = - \mat B(\vec y) \vec \nabla_{\vec y} {H} .
\end{equation}
The so-called Poisson matrix $\mat B(\vec y)$ is
\begin{equation}
\mat B(\vec y) = \begin{bmatrix}
\mat C & \mat X \\
-\trans{\mat X} & \mat 0
\end{bmatrix}
\end{equation}
where $\trans{(\cdot)}$ means the transpose of a vector or of a matrix.  $\mat X$ is an $n\times m$
matrix and $\mat C$ an $n\times n$ matrix whose elements are
\begin{equation}
[\mat X]_{ij} = X_{ij}
\qquad\mathrm{and}\qquad
[\mat C]_{ij} = -\sum_k c_{ij}^k \pi_k.
\end{equation}

\subsubsection{Translation group}
The simplest illustration of the above formalism is the case where $G$ is the translation group. In
that case, $\vec \eta$ is the usual velocity vector $\vec v$ and $\vec \pi$ is the standard linear
momentum, commonly denoted $\vec p$. The vector fields of the tangent configuration space are
$\field X_i = \frac{\partial}{\partial{q_i}}$. The associated structure constants $c_{ij}^k$ are all
nil. The Poisson matrix is then
\begin{equation}
\mat B(\vec y) = \begin{bmatrix}
\mat 0 & \mat 1 \\
-\mat 1 & \mat 0
\end{bmatrix}
\end{equation}
and we retrieve the canonical equations of motion
\begin{equation}
\dot p_i = -\frac{\partial {H}}{\partial q_i} ,
\qquad
\dot q_i = \frac{\partial {H}}{\partial p_i}\ .
\end{equation}

\subsubsection{Group SO(3) in the body-fixed frame}
The group SO(3) naturally appears in studies of the rotation motion of solid bodies. For this
problem, two choices can be made: vectors are expressed either in the body-fixed frame or in the
``laboratory'' frame.  Here, we consider the first option where vectors are written in the
body-fixed frame. The Lie velocity is the rotation vector designated by $\vec \omega$ and the
orientation of the body is parametrised by the coordinates in the body-fixed frame of the laboratory
base vectors, i.e., $\vec q = (\vec i, \vec j, \vec k)$. For any function $f(\vec i, \vec j, \vec
k)$, we have
\begin{equation}
\begin{split}
\frac{\dd}{\dd t}f(\vec i, \vec j, \vec k) &= 
 -(\vec \omega \times \vec i)\cdot \frac{\partial f}{\partial \vec i}
 -(\vec \omega \times \vec j)\cdot \frac{\partial f}{\partial \vec j}
 -(\vec \omega \times \vec k)\cdot \frac{\partial f}{\partial \vec k} \\
 &=
 -\vec \omega \cdot \left(
    \vec i\times \frac{\partial f}{\partial \vec i}
   +\vec j\times \frac{\partial f}{\partial \vec j}
   +\vec k\times \frac{\partial f}{\partial \vec k}
\right) .
\end{split}
\end{equation}
Thus, the vector field ${\vfield X} = (\field X_1, \field X_2, \field X_3)$ is
\begin{equation}
{\vfield X} =
          - \vec i\times \frac{\partial}{\partial \vec i}
          - \vec j\times \frac{\partial}{\partial \vec j}
          - \vec k\times \frac{\partial}{\partial \vec k}\ ,
\end{equation}
with structure constants $c_{ij}^k = -\epsilon_{ijk}$ where $\epsilon_{ijk}=1$ when $(i,j,k)$ is a
cyclic permutation of $(1,2,3)$, -1 when $(i,j,k)$ is a cyclic permutation of $(3,2,1)$, 0
otherwise.  Hence, the Poisson matrix reads
\begin{equation}
\mat B = - \begin{bmatrix}
\hvec \pi & \hvec \imath & \hvec \jmath & \hvec k \\
\hvec \imath & \mat 0 & \mat 0 & \mat 0 \\
\hvec \jmath & \mat 0 & \mat 0 & \mat 0 \\
\hvec k      & \mat 0 & \mat 0 & \mat 0
\end{bmatrix}
\end{equation}
where for any vector $\vec v$, we have defined
\begin{equation}
\hvec v = \begin{bmatrix}
0 & -v_z & v_y \\
v_z & 0 & -v_x \\
-v_y & v_x & 0
\end{bmatrix}\ .
\end{equation}
The corresponding equations of motion are
\begin{eqnarray}
\frac{\dd\vec \pi}{\dd t} &=& 
  \vec \pi \times \frac{\partial {H}}{\partial \vec \pi}
+ \vec i \times \frac{\partial {H}}{\partial \vec i}
+ \vec j \times \frac{\partial {H}}{\partial \vec j}
+ \vec k \times \frac{\partial {H}}{\partial \vec k} , \\
\frac{\dd\vec i}{\dd t} &=& \vec i \times \frac{\partial {H}}{\partial \vec \pi} , \\
\frac{\dd\vec j}{\dd t} &=& \vec j \times \frac{\partial {H}}{\partial \vec \pi} , \\
\frac{\dd\vec k}{\dd t} &=& \vec k \times \frac{\partial {H}}{\partial \vec \pi}
\end{eqnarray}
with $\partial{H}/\partial \vec \pi = \vec \omega$.

\subsubsection{Group SO(3) in the laboratory frame}
\label{sec.SO3inertial}
Here we again consider the rotation motion of a solid body but now vector coordinates are written in
the laboratory frame. The latter is the frame with respect to which the motion of the
spinning body is described. Note that it does not have to be inertial. The generalised coordinates
are the base vectors of the rotated frame $\vec q = (\vec I, \vec J, \vec K)$ and the Lie momentum
associated with the rotation vector is denoted $\vec \Pi$. Applying the same method as
above, we get
\begin{equation}
{\vfield X} = \vec I \times \frac{\partial }{\partial \vec I}
        + \vec J \times \frac{\partial }{\partial \vec J}
        + \vec K \times \frac{\partial }{\partial \vec K} .
\end{equation}
For this basis, the structure constants are $c_{ij}^k = \epsilon_{ijk}$ and thus, the Poisson matrix
is
\begin{equation}
\mat B = \begin{bmatrix}
\hvec \Pi & \hvec I & \hvec J & \hvec K \\
\hvec I & \mat 0 & \mat 0 & \mat 0 \\
\hvec J & \mat 0 & \mat 0 & \mat 0 \\
\hvec K & \mat 0 & \mat 0 & \mat 0
\end{bmatrix}\ .
\label{eq.BY}
\end{equation}
The associated equations of motion are
\begin{eqnarray}
\frac{\dd\vec \Pi}{\dd t} &=& 
  \frac{\partial {H}}{\partial \vec \Pi} \times
\vec \Pi
+ \frac{\partial {H}}{\partial \vec I} \times \vec I
+ \frac{\partial {H}}{\partial \vec J} \times \vec J
+ \frac{\partial {H}}{\partial \vec K} \times \vec K , \\
\frac{\dd\vec I}{\dd t} &=& \frac{\partial {H}}{\partial \vec \Pi} \times \vec I , \\
\frac{\dd\vec J}{\dd t} &=& \frac{\partial {H}}{\partial \vec \Pi} \times \vec J , \\
\frac{\dd\vec K}{\dd t} &=& \frac{\partial {H}}{\partial \vec \Pi} \times \vec K
\end{eqnarray}
where $\partial {H}/\partial \vec \Pi$ still is the rotation vector, although expressed in the
laboratory frame.

\subsection{Linearisation and driven solution}
\label{sec.linearisation}
For the sake of completeness, we here recall the general method leading to the linearisation of the
equations of motion in the non-canonical Hamiltonian formalism \citep{Maddocks91,Beck98}. We also
present the criterion of nonlinear stability as described in {\em ibid.}

Let a non-autonomous Hamiltonian $H(\vec y, t)$ associated with an $n$ degrees of freedom
system expressed as a function of non-canonical variables $\vec y \in \mathbb{R}^p$ with $p\geq 2n$.
We assume that $H(\vec y, t)$ can be split as follows
\begin{equation}
H(\vec y, t) = H_0(\vec y) + H_1(\vec y, t) ,
\end{equation}
where $H_0(\vec y)$ is the autonomous part of $H(\vec y, t)$ and $H_1(\vec y, t)$ a small
perturbation. Let us skip the perturbation $H_1$ for a moment. The equations of motion
associated with $H_0(\vec y)$ are of the form
\begin{equation}
\dot{\vec y} = - \mat B(\vec y) \vec \nabla_{\vec y} H_0(\vec y) .
\label{eq.doty}
\end{equation}

The system has $n$ degrees of freedom, its phase space $\Sigma$ is thus a manifold of dimension
$2n$. Since $\vec y\in \mathbb{R}^p$, there exists $s=p-2n$ Casimir functions $C_i(\vec y)$ and $s$
constants $c_i$, $1\leq i \leq s$, such that
\begin{equation}
\Sigma = \{ \vec y \in \mathbb{R}^p \colon C_1(\vec y) = c_1, \ldots,
C_s(\vec y) = c_s \} .
\end{equation}
We recall that Casimir functions are constants of the motion for any Hamiltonian because their
gradients constitute a basis of the kernel of the Poisson matrix:
\begin{equation}
\ker \mat B(\vec y) = \mathrm{span} \left\{
\vec \nabla_{\vec y} C_1(\vec y), \ldots, \vec \nabla_{\vec y} C_s(\vec
y)
\right\} ,
\end{equation}
and thus
\begin{equation}
\dot C_i(\vec y) = \trans{(\vec \nabla_{\vec y} H_0)}
\mat B(\vec y) \vec \nabla_{\vec y} C_i = 0
\label{eq.dotCi}
\end{equation}
for all Hamiltonian $H_0$.

Let $\vec y_e$ be an equilibrium, i.e., a fixed point of $H_0$. According to
Eq.~(\ref{eq.doty}), $\dot{\vec y}_e = \vec 0$ implies $\vec \nabla_{\vec y} H_0(\vec y_e) \in \ker
\mat B(\vec y_e)$. Thus, there exists $s$ coefficients $(\mu_i)_{1\leq i \leq s}$ such that
\begin{equation}
\vec \nabla_{\vec y} H_0(\vec y_e) = \sum_{i=1}^s \mu_i \vec
\nabla_{\vec y} C_i(\vec y_e) .
\end{equation}
Let
\begin{equation}
F(\vec y) = H_0(\vec y) - \sum_{i=1}^s \mu_i C_i(\vec y) .
\end{equation}
By construction, $F$ satisfies $\vec \nabla_{\vec y}F(\vec y_e) = \vec 0$. Coefficients $\mu_i$ can
be seen as Lagrange multipliers and functions $C_i(\vec y)$ as constraints since  we search for an
extremum of $H_0(\vec y)$ under the conditions $C_i(\vec y) = c_i$.  The $p+s$ equations $\vec
\nabla_{\vec y}F(\vec y_e) = \vec 0$ and $C_i(\vec y_e) = c_i$ allow to determine $\vec y_e$ and the
coefficients $\mu_i$.

Once $\vec y_e$ and coefficients $\mu_i$ are known, the linearisation of the equations of motion
(Eq.~\ref{eq.doty}) are given by
\begin{equation}
\delta \dot{\vec y} =  \mat A(\vec y_e) \delta \vec y
\end{equation}
with $\delta {\vec y} = \vec y - \vec y_e$ and \citep{Maddocks91}
\begin{equation}
\mat A(\vec y_e) = -\mat B(\vec y_e) \mat \nabla^2_{\vec y} F(\vec y_e) .
\label{eq.matA}
\end{equation}
In a last step, the perturbation $H_1(\vec y, t)$ is taken into account and the equations of motion
become
\begin{equation}
\delta \dot{\vec y}  - \mat A(\vec y_e) \delta \vec y = \vec z(t) ,
\label{eq.driven}
\end{equation}
with
\begin{equation}
\vec z(t) = -\mat B(\vec y_e) \vec \nabla_{\vec y} H_1(\vec y_e, t) .
\end{equation}
Equation~(\ref{eq.driven}) is then solved using standard techniques.

The relative equilibria $\vec y = \vec y_e$ is said to be nonlinearly stable if the quadratic form
(or Lyapunov function) $N(\vec y) = \trans{\vec y} \mat N \vec y$, defined on the phase space
$\Sigma$ by its Hessian (below), is a strictly convex function \citep{Beck98}. The Hessian of
$N(\vec y)$ is given by (see {\em ibid.})
\begin{equation}
\mat N \equiv \nabla^2 N = \mat Q(\vec y_e) \nabla^2 F(\vec y_e) \mat
Q(\vec y_e) ,
\label{eq.Lyap}
\end{equation}
where $\mat Q(\vec y)$ is the orthogonal projection matrix onto the range of
$\mat A(\vec y)$,
\begin{equation}
\mat Q(\vec y) = \mat 1 - \mat K(\vec y)(\trans{\mat K}(\vec y)\mat
K(\vec y))^{-1}\trans{\mat K}(\vec y) ,
\end{equation}
and where $\mat K(\vec y)$ is a $p\times s$ matrix given by
\begin{equation}
\mat K(\vec y) = \begin{bmatrix} \vec \nabla C_1(\vec y) & \cdots & \vec
\nabla C_s(\vec y)\end{bmatrix} .
\end{equation}

\tabnotation

\figmodels

\section{Rigid satellite}
\label{sec.rigid-satellite}
Let a rigid satellite whose rotation is close to the synchronous state, i.e., whose mean rotation
rate is equal to the orbital mean motion.  The goal of this section is to compute the frequencies
associated with the free modes of rotation, to evaluate the forced obliquity driven by the
orbital precession, and eventually to check the nonlinear stability of the system in the vicinity of
the equilibrium. The analysis is performed using the non-canonical Hamiltonian formalism described
in Sect.~\ref{sec.non-canonical}. It turns out to be convenient to describe the problem in a
laboratory frame rotating at constant angular speed $\vec \Omega$ with respect to the inertial
frame.  $\Omega$ will then be chosen equal to the mean orbital motion. We denote by $\vec \omega$
the rotation vector of the satellite with respect to the laboratory frame $\mathcal{F}_\mathrm{lab}$
and by $(\vec I, \vec J, \vec K)$ its principal axes of inertia such that the matrix of inertia
reads
\begin{equation}
\inertia = \mat R \diag(A, B, C) \trans{\mat R} ,
\end{equation}
where $\mat R = [\vec I, \vec J, \vec K]$ is the rotation matrix of the satellite with respect to
the laboratory frame and where $\trans{(.)}$ denotes the transpose operator. Note that the matrix of
inertia can also be written in a equivalent form facilitating the computation of the gradient of the
forthcoming Hamiltonian
\begin{equation}
\inertia  = A\vec I\trans{\vec I}
          + B\vec J\trans{\vec J}
          + C\vec K\trans{\vec K} .
\end{equation}
The Lie velocity of the system is thus $\vec \omega$ while $(\vec I, \vec J, \vec K)$ are the
generalised coordinates. We also denote by $(\vec i, \vec j, \vec k)$ the basis vectors
associated with the laboratory frame. The radius vector connecting the planet and the
satellite barycenter is assumed to be a known function of time and is denoted either by $\vec r(t)$
or simply by $\vec r$. $\mathcal{G}$ and $M_p$ are the gravitational constant and the mass of the
planet, respectively. With these notations, the (non-autonomous) Lagrangian $L_\mathrm{rs}(\vec
\omega,\vec I, \vec J, \vec K,t)$ governing the rotation of the rigid satellite is
\begin{equation}
L_\mathrm{rs}(\vec \omega, \vec I, \vec J, \vec K, t) = 
\frac{\trans{(\vec \omega + \vec \Omega)} \inertia (\vec \omega + \vec
\Omega)}{2} - \frac{3{\cal G}M_p}{2} \frac{\trans{\vec r} \inertia \vec r}{r^5}\ .
\end{equation}
The Lie momentum $\vec \Pi$ associated with $\vec \omega$ reads
\begin{equation}
\vec \Pi = \frac{\partial L_\mathrm{rs}}{\partial \vec \omega} = \inertia (\vec
\omega + \vec \Omega) .
\end{equation}
We recognise the spin angular momentum of the satellite with respect to the inertial frame and
expressed in the laboratory frame.  The Hamiltonian $H_\mathrm{rs}(\vec \Pi, \vec I, \vec J, \vec K,
t)$ resulting from the Legendre transformation applied to $L_\mathrm{rs}(\vec \omega, \vec I, \vec
J, \vec K, t)$ reads
\begin{equation}
H_\mathrm{rs}(\vec \Pi, \vec I, \vec J, \vec K, t) = \frac{\trans{\vec \Pi} 
\inertia^{-1} \vec \Pi}{2} - \trans{\vec \Omega} \vec \Pi 
+ \frac{3{\cal G}M_p}{2} \frac{\trans{\vec r} \inertia \vec r}{r^5}
\end{equation}
with
\begin{equation}
\inertia^{-1} = \frac{\vec I\trans{\vec I}}{A}
              + \frac{\vec J\trans{\vec J}}{B}
              + \frac{\vec K\trans{\vec K}}{C} .
\end{equation}
The Poisson matrix $\mat B_\mathrm{rs}(\vec y)$ associated with $\vec y = (\vec \Pi, \vec I,
\vec J, \vec K)$ is the one given in Eq.~(\ref{eq.BY}). The gradient of the Hamiltonian reads
\begin{eqnarray}
&& \frac{\partial H_\mathrm{rs}}{\partial \vec \Pi} = \inertia^{-1} \vec \Pi - \vec
\Omega = \vec \omega , \\
&& \frac{\partial H_\mathrm{rs}}{\partial \vec I} = 
  \frac{(\vec I\cdot \vec \Pi)}{A}\vec \Pi
+ 3\frac{{\cal G}M_p}{r^5} A(\vec r \cdot \vec I)\vec r , \\
&& \frac{\partial H_\mathrm{rs}}{\partial \vec J} = 
  \frac{(\vec J\cdot \vec \Pi)}{B}\vec \Pi
+ 3\frac{{\cal G}M_p}{r^5} B(\vec r \cdot \vec J)\vec r , \\
&& \frac{\partial H_\mathrm{rs}}{\partial \vec K} = 
  \frac{(\vec K\cdot \vec \Pi)}{C}\vec \Pi
+ 3\frac{{\cal G}M_p}{r^5} C(\vec r \cdot \vec K)\vec r ,
\end{eqnarray}
and thus the equations of motion are
\begin{eqnarray}
\label{eq.dotPi}
\dot{\vec \Pi} &=& \vec \Pi \times \vec \Omega 
               - 3\frac{{\cal G} M_p}{r^5} (\inertia \vec r)\times \vec r ,\\
\label{eq.dotI}
\dot{\vec I} &=& \vec \omega \times \vec I , \\
\label{eq.dotJ}
\dot{\vec J} &=& \vec \omega \times \vec J , \\
\label{eq.dotK}
\dot{\vec K} &=& \vec \omega \times \vec K .
\end{eqnarray}%

Equations of motion (Eqs.~\ref{eq.dotPi}-\ref{eq.dotK}) are those of the full Hamiltonian. Because
$\vec r(t)$ is a function of time, the set of equations~(\ref{eq.dotPi}-\ref{eq.dotK}) has no fixed
point. To proceed, we set $\vec \Omega = \Omega \vec k$ with $\Omega$ equal to the mean orbital
motion such that, in the laboratory frame $(\vec i, \vec j, \vec k)$,
\begin{equation}
\mat S(t) \equiv
{\cal G}M_p \frac{\vec r\trans{\vec r}}{r^5} = \mat S_0 +
\mat S_1(t)
\label{eq.matS}
\end{equation}
where $\mat S_0$ is a constant matrix and $\mat S_1(t)$ a small perturbation. Furthermore, the
initial angle of the rotation is chosen such that $\mat S_0$ is diagonal with components
$(\sigma^0_{xx}, \sigma^0_{yy}, \sigma^0_{zz})$. Similarly, we denote by $\sigma^1_{uv}$, where $u,v
\in \{x,y,z\}$, the elements of $\mat S_1(t)$. The gravitational potential energy $U(\vec y, t)$ is
then split into $U_0(\vec y) + U_1(\vec y, t)$ with
\begin{eqnarray}
\label{eq.U0}
&& U_0(\vec y) = 
      \frac{3}{2}\left(A\trans{\vec I} \mat S_0 \vec I
                     + B\trans{\vec J} \mat S_0 \vec J
                     + C\trans{\vec K} \mat S_0 \vec K\right) , \\
\label{eq.U1}
&& U_1(\vec y, t) = \frac{3}{2}\left(A\trans{\vec I} \mat S_1(t) \vec I
                       + B\trans{\vec J} \mat S_1(t) \vec J
                       + C\trans{\vec K} \mat S_1(t) \vec K\right) .
\end{eqnarray}
As a result, the Hamiltonian $H_\mathrm{rs}(\vec y, t)$ also get split into $H_{\mathrm{rs}}^0(\vec
y) + H_{\mathrm{rs}}^1(\vec y, t)$ with
\begin{eqnarray}
H_{\mathrm{rs}}^0(\vec y) &=& \frac{\trans{\vec \Pi} \inertia^{-1}\vec \Pi}{2}
    - \trans{\vec \Omega} \vec \Pi + U_0(\vec y) , \\
H_{\mathrm{rs}}^1(\vec y,t) &=& U_1(\vec y, t) .
\end{eqnarray}
In the case of a Keplerian orbit with eccentricity $e$ and inclination $i$ with respect to the
reference frame,
\begin{eqnarray}
&&\sigma^0_{xx} = \frac{{\cal G}M_p}{a^3}\left(\frac{X_0^{-3,0}(e)+X_2^{-3,2}(e)}{2} \cos^4\left(\frac{i}{2}\right)
                + \frac{X_0^{-3,0}(e)}{2}               \sin^4\left(\frac{i}{2}\right)\right) , \\
&&\sigma^0_{yy} = \frac{{\cal G}M_p}{a^3}\left(\frac{X_0^{-3,0}(e)-X_2^{-3,2}(e)}{2} \cos^4\left(\frac{i}{2}\right)
                + \frac{X_0^{-3,0}(e)}{2}               \sin^4\left(\frac{i}{2}\right)\right) , \\
&&\sigma^0_{zz} = \frac{{\cal G}M_p}{a^3}\frac{X_0^{-3,0}(e)}{2} \sin^2 i ,
\end{eqnarray}
where $X_k^{n,m}(e)$ are Hansen coefficients \citep{Hansen55} defined as Fourier coefficients of the
series
\begin{equation}
\left(\frac{r}{a}\right)^n \e^{\ii m v} = \sum_{k=-\infty}^\infty
X_k^{n,m}(e) \e^{\ii k M}
\end{equation}
with $a$, $v$, $M$ being the semimajor axis, the true anomaly and the mean anomaly, respectively.
Besides, in this study a single element of the matrix $\mat S_1(t)$ plays a role in
the tilting of the Cassini state, this is the term in $\sigma^1_{xz}(t) = \sigma^1_{zx}(t)$
corresponding to the first harmonic of the orbital precession in inclination whose expression is
\begin{equation}
\sigma^1_{xz}(t) = \frac{{\cal G}M_p}{a^3}
 \left(\frac{X_0^{-3,0}(e)}{2}\cos i
      +\frac{X_2^{-3,2}(e)}{2}\cos^2\frac{i}{2}\right)\sin i\sin(\Omega t - \Phi)
\label{eq.sig1}
\end{equation}
where $\Phi$ is the longitude of the ascending node. The expression of the Hansen coefficients
involved in $\mat S_0$ and
$\mat S_1(t)$ are
\begin{eqnarray}
&&X_0^{-3,0}(e) = (1-e^2)^{-3/2} , \\
&&X_2^{-3,2}(e) = 1 - \frac{5}{2}e^2 + \frac{13}{16}e^4 - \frac{35}{288}e^6 + O(e^8) .
\end{eqnarray}

Following the steps recalled in the previous section~\ref{sec.linearisation}, we now skip the
perturbation $\mat S_1(t)$ for a while and only retain the autonomous part of the Hamiltonian
$H_\mathrm{rs}^0(\vec y)$.  The gradient of the Hamiltonian $H_\mathrm{rs}^0(\vec y)$ reads 
\begin{eqnarray}
\frac{\partial H_\mathrm{rs}^0}{\partial \vec \Pi} &=& \inertia^{-1}\vec \Pi - \vec
\Omega = \vec \omega , \\
\frac{\partial H_\mathrm{rs}^0}{\partial \vec I} &=& \frac{(\vec I\cdot\vec \Pi)}{A}\vec \Pi + 3 A \mat S_0 \vec I , \\
\frac{\partial H_\mathrm{rs}^0}{\partial \vec J} &=& \frac{(\vec J\cdot\vec \Pi)}{B}\vec \Pi + 3 B \mat S_0 \vec J , \\
\frac{\partial H_\mathrm{rs}^0}{\partial \vec K} &=& \frac{(\vec K\cdot\vec \Pi)}{C}\vec \Pi + 3 C \mat S_0 \vec K .
\end{eqnarray}
Only $\dot{\vec \Pi}$ (Eq.~\ref{eq.dotPi}) is affected by the averaging process. Its new equation of
motion reads
\begin{equation}
\dot{\vec \Pi} = \vec \Pi \times \vec \Omega 
                + 3A(\mat S_0 \vec I)\times \vec I
                + 3B(\vec S_0 \vec J)\times \vec J
                + 3C(\vec S_0 \vec K)\times \vec K .
\label{eq.dotPi0}
\end{equation}

\subsection{Linearisation}
To perform the linearisation of Eqs.~(\ref{eq.dotPi0}, \ref{eq.dotI}-\ref{eq.dotK}), we note that
the phase space $\Sigma_\mathrm{rs}$ of the system is a manifold of dimension 6 (associated
with the 3 degrees of freedom of the group SO(3)) defined as
\begin{equation}
\begin{split}
\Sigma_\mathrm{rs} = \{\vec y\in \mathbb{R}^{12} \colon 
  C_\mathrm{rs}^1(\vec y) = C_\mathrm{rs}^2(\vec y) = C_\mathrm{rs}^3(\vec y) = 1/2, \\
  C_\mathrm{rs}^4(\vec y) = C_\mathrm{rs}^5(\vec y) = C_\mathrm{rs}^6(\vec y) = 0\} ,
\end{split}
\end{equation}
where the Casimir functions are
\begin{align}
&
C_\mathrm{rs}^1(\vec y) = \frac{1}{2}\vec I\cdot \vec I, \quad
C_\mathrm{rs}^2(\vec y) = \frac{1}{2}\vec J\cdot \vec J, \quad
C_\mathrm{rs}^3(\vec y) = \frac{1}{2}\vec K\cdot \vec K, \nonumber \\
&
C_\mathrm{rs}^4(\vec y) =            \vec J\cdot \vec K, \quad
C_\mathrm{rs}^5(\vec y) =            \vec K\cdot \vec I, \quad
C_\mathrm{rs}^6(\vec y) =            \vec I\cdot \vec J .
\end{align}
Indeed, it can be checked that
\begin{equation}
\ker \mat B_\mathrm{rs}(\vec y) = \mathrm{span} \left\{ 
\begin{pmatrix}
\vec 0 \\ \vec I \\ \vec 0 \\ \vec 0
\end{pmatrix},
\begin{pmatrix}
\vec 0 \\ \vec 0 \\ \vec J \\ \vec 0
\end{pmatrix},
\begin{pmatrix}
\vec 0 \\ \vec 0 \\ \vec 0 \\ \vec K
\end{pmatrix},
\begin{pmatrix}
\vec 0 \\ \vec 0 \\ \vec K \\ \vec J
\end{pmatrix},
\begin{pmatrix}
\vec 0 \\ \vec K \\ \vec 0 \\ \vec I
\end{pmatrix},
\begin{pmatrix}
\vec 0 \\ \vec J \\ \vec I \\ \vec 0
\end{pmatrix}
\right\} .
\end{equation}
Let $F_\mathrm{rs}(\vec y) = H_\mathrm{rs}^0(\vec y) - \sum_i \mu_i C_\mathrm{rs}^i(\vec y)$. The
condition $\vec \nabla_{\vec y} F_\mathrm{rs}(\vec y_e) = \vec 0$ leads to
\begin{eqnarray}
\inertia^{-1} \vec \Pi_e - \vec \Omega = \vec \omega_e &=& \vec 0 , \\
\frac{(\vec I_e\cdot\vec \Pi_e)}{A}\vec \Pi_e + 3A\mat S_0 \vec I_e 
    - \mu_1 \vec I_e - \mu_5 \vec K_e - \mu_6 \vec J_e &=& \vec 0 , \\
\frac{(\vec J_e\cdot\vec \Pi_e)}{B}\vec \Pi_e + 3B\mat S_0 \vec J_e 
    - \mu_2 \vec J_e - \mu_4 \vec K_e - \mu_6 \vec I_e &=& \vec 0 , \\
\frac{(\vec K_e\cdot\vec \Pi_e)}{C}\vec \Pi_e + 3C\mat S_0 \vec K_e 
    - \mu_3 \vec K_e - \mu_4 \vec J_e - \mu_5 \vec I_e &=& \vec 0 ,
\end{eqnarray}
whose a solution is
\begin{align}
\vec \omega_e = \vec 0 ,\quad
\vec \Pi_e = C\Omega \vec k ,\quad
\vec I_e = \vec i ,\quad
\vec J_e = \vec j ,\quad
\vec K_e = \vec k , \nonumber \\
\mu_1 = 3A\sigma^0_{xx}, \quad
\mu_2 = 3B\sigma^0_{yy}, \quad
\mu_3 = 3C\sigma^0_{zz} + C\Omega^2 ,\quad
\mu_4 = \mu_5 = \mu_6 = 0 .
\end{align}
The other solutions are equivalent to this one but with a permutation of the moments of inertia $A$,
$B$, $C$. The matrix $\mat A_\mathrm{rs}(\vec y_e)$ of the linearised system is given by
Eq.~(\ref{eq.matA}). To simplify the result, we perform the change of variables $\delta{\vec y} =
\mat P
\delta{\vec y}^*$ with
\begin{equation}
\begin{split}
\delta{\vec y}^* = \Big(\begin{array}{*{9}c}
\delta \Pi_z, & 
\delta I_y, & 
\delta \Pi_x, & 
\delta \Pi_y, & 
\delta I_z, & 
\delta J_z, & 
\delta I_x, & 
\delta J_y, & 
\delta K_z,  
\end{array} \\
\trans{
\begin{array}{*{3}c}
\delta I_y + \delta J_x, &
\delta I_z + \delta K_x, & 
\delta J_z + \delta K_y
\end{array}\Big)} .
\end{split}
\label{eq.deltay*}
\end{equation}
The first two components of $\delta{\vec y}^*$ are associated with the libration in
longitude, the next four components describe the wobble and the libration in latitude, and finally,
the last six coordinates being in the kernel of $\mat B_\mathrm{rs}(\vec y_e)$ remain identically
equal to zero.  Let $\mat A_{\mathrm{rs}}^*(\vec y_e)$ be the matrix of the linear system in the new
variables $\delta{\vec y}^*$, i.e., $\mat A_{\mathrm{rs}}^* = \mat P^{-1} \mat A_\mathrm{rs} \mat
P$, and let $\mat A_\mathrm{rs}^1$ and $\mat A_\mathrm{rs}^2$ be the respective $2\times2$ and
$4\times4$ matrices such that
\begin{equation}
\mat A_\mathrm{rs}^*(\vec y_e) = \begin{bmatrix}
\mat A_\mathrm{rs}^1 & \mat 0 & \cdot  \\
\mat 0 & \mat A_\mathrm{rs}^2 & \cdot  \\
\mat 0 & \mat 0 & \mat 0
\end{bmatrix}
\end{equation}
where the dots $\cdot$ represent arbitrary matrices not influencing the motion. We have
\begin{equation}
\mat A_\mathrm{rs}^1 = \begin{bmatrix}
0 & -3(B-A)(\sigma^0_{xx}-\sigma^0_{yy}) \\
1/C & 0 
\end{bmatrix} ,
\end{equation}
and
\begin{equation}
\mat A_\mathrm{rs}^2 = \begin{bmatrix}
0 & \Omega & 0 & 3(C-B)(\sigma^0_{zz}-\sigma^0_{yy}) \\
-\Omega & 0 & 3(C-A)(\sigma^0_{xx}-\sigma^0_{zz}) & 0 \\
0 & -\Frac{1}{B} & 0 & -\Frac{C-B}{B}\Omega \\
\Frac{1}{A} & 0 & \Frac{C-A}{A}\Omega & 0 
\end{bmatrix} .
\end{equation}
Hence, the frequency of libration in longitude $\omega_{\mathrm{rs},u}$, which is the eigenvalue
of $\mat A_\mathrm{rs}^1$, reads
\begin{equation}
\omega_{\mathrm{rs},u} = \sqrt{3\gamma\left(\kappa_1-\kappa_2\right)} ,
\label{eq.nulong}
\end{equation}
and the frequencies associated with the wobble $\omega_{\mathrm{rs},w}$ and the libration in
latitude $\omega_{\mathrm{rs},v}$, the eigenvalues of $\mat A_\mathrm{rs}^2$, are given by
\begin{equation}
\omega_{\mathrm{rs},w} = \left(\frac{p-\sqrt{p^2-4q}}{2}\right)^{1/2} ,
\qquad
\omega_{\mathrm{rs},v} = \left(\frac{p+\sqrt{p^2-4q}}{2}\right)^{1/2}
\label{eq.nuwl}
\end{equation}
with
\begin{eqnarray}
&&p = \left(1+\alpha\beta\right)\Omega^2 + 3\left(\beta \kappa_1 + \alpha \kappa_2\right) , \\
&&q = \alpha\beta\left(\Omega^4 + 3\left(\kappa_1+\kappa_2\right)\Omega^2 + 9 \kappa_1\kappa_2\right) , \\
&&\kappa_1 = \sigma^0_{xx}-\sigma^0_{zz} , \\
&&\kappa_2 = \sigma^0_{yy}-\sigma^0_{zz} ,
\end{eqnarray}
and
\begin{equation}
\alpha = \frac{C-B}{A} ,
\qquad
\beta = \frac{C-A}{B} ,
\qquad
\gamma = \frac{B-A}{C} .
\end{equation}
Here we retrieve the well-known eigenfrequencies of a rigid satellite close to the synchronous
equilibrium state \citep[e.g.,][]{Rambaux12}.  Let us nevertheless stress that
Eqs.~(\ref{eq.nulong}) and (\ref{eq.nuwl}) are associated with the motion of the {\em three}
vectors $(\vec I, \vec J, \vec K)$ in the rotating frame. By consequence, if we denote by
$\bar{\omega}_{\mathrm{rs},v} \approx 3\beta\Omega/2$ the frequency of libration in latitude
associated with the motion of the {\em sole} vector $\vec K$ with respect to the inertial
frame (as it is commonly defined for an axisymmetric body), we have $\omega_{\mathrm{rs},v} =
\bar{\omega}_{\mathrm{rs},v} + \Omega$.

\subsection{Stability}
For this problem, the Lyapunov function $N_\mathrm{rs}(\vec y)$, as defined in Eq.~(\ref{eq.Lyap}),
is
\begin{equation}
\begin{split}
N_\mathrm{rs}(\vec y) = &
  \frac{1}{2A}\left(\Pi_x+\frac{1}{2}(C-A)\Omega(I_z-K_x)\right)^2
+ \frac{1}{2B}\left(\Pi_y+\frac{1}{2}(C-B)\Omega(J_z-K_y)\right)^2
\\ &
+ \frac{1}{2C}\Pi_z^2
+ \frac{1}{2}n_1(I_y-J_x)^2
+ \frac{1}{2}n_2(I_z-K_x)^2
+ \frac{1}{2}n_3(J_z-K_y)^2
\end{split}
\end{equation}
with
\begin{equation}
n_1 = \frac{3}{4}(B-A)(\kappa_1-\kappa_2) ,
\quad
n_2 = \frac{1}{8}(C-A)(\Omega^2+3\kappa_1) ,
\quad
n_3 = \frac{1}{8}(C-B)(\Omega^2+3\kappa_2) .
\label{eq.n123}
\end{equation}
We recall that the system is nonlinearly stable if $N_\mathrm{rs}(\vec y)$ is a strictly convex
function. Coefficients $A$, $B$, and $C$ are positive, as required. The nonlinear stability is then
achieved when $n_1$, $n_2$, and $n_3$ are all positive. Given that $\kappa_1> \kappa_2
> 0$ at low inclination $i$, the criterion implies $C>B>A$, which is the well-known stability
condition for this classical equilibrium where the longest axis points towards the parent planet
\citep[e.g.][]{Beck98}.

\subsection{Driven solution}
Here we look for the forced solution when the time-dependent perturbation
$H_{\mathrm{rs}}^1(t)$ is taken into account. In the variables $\delta \vec y^*$
(Eq.~\ref{eq.deltay*}), and with the notation of Eq.~(\ref{eq.driven}), the perturbation
$\delta \vec z_\mathrm{rs}^*(t)$ is given by
\begin{equation}
\delta \vec z_\mathrm{rs}^*(t) = -\mat P^{-1} \mat B_\mathrm{rs}(\vec y_e) \vec \nabla_{\vec
y}H_\mathrm{rs}^1(\vec y_e, t) .
\label{eq.deltazrs}
\end{equation}
To match the notation of the matrix $\mat A_\mathrm{rs}^*$, let $\delta
\vec y^1$ and $\delta \vec y^2$ be the first 2 and the next 4 components of
$\delta \vec y^*$, idem for $\delta \vec z_\mathrm{rs}^*(t)$, such that
the linear problem with perturbation reads
\begin{eqnarray}
\delta \dot{\vec y}^k - \mat A_\mathrm{rs}^k \delta \vec y^k = \delta
\vec z_\mathrm{rs}^k(t) , \qquad k=1,2 .
\end{eqnarray}
By definition,
\begin{equation}
\delta \vec y^1 = \trans{(\delta \Pi_z, \delta I_y)} , \qquad
\delta \vec y^2 = \trans{(\delta \Pi_x, \delta \Pi_y, \delta I_z, \delta J_z)} ,
\end{equation}
and Eq.~(\ref{eq.deltazrs}) implies
\begin{equation}
\delta \vec z_\mathrm{rs}^1(t) = \begin{pmatrix}
3 (B-A)\sigma^1_{xy}(t) \\
0 \end{pmatrix} , \qquad
\delta \vec z_\mathrm{rs}^2(t) = \begin{pmatrix}
3(C-B)\sigma^1_{yz}(t) \\
 -3(C-A)\sigma^1_{xz}(t) \\ 0 \\ 0
\end{pmatrix} .
\end{equation}
Note that the term $\sigma_{yz}^1(t)$ is present in the perturbation $\delta \vec
z_\mathrm{rs}^2(t)$ but its effect on the orientation of the spin axis is very weak. For instance,
according to the ephemeris of Titan in TASS1.6 \citep{Vienne95}, the amplitude associated with the angle
$(\Omega t-\Phi)$ in $\sigma_{yz}^1(t)$ is about 500 times lower than that in $\sigma_{xz}^1(t)$. In
the numerical applications (Sect.~\ref{sec.application}), $\sigma_{yz}^1(t)$ is simply discarded.

\section{Satellite with a liquid core}
\label{sec.liquid-core}
In this section we consider a satellite with a rigid mantle/crust layer surrounding a liquid core.
In a first step, we analyse the problem using the Poincar\'e-Hough model which is valid for all
eccentricities of the ellipsoidal cavity containing the fluid core \citep{Poincare10, Hough95}.  In
a second one, we truncate the problem at the first order with respect to the equatorial and polar
flattening of the cavity. The same simplification will be used again in Sect.~\ref{sec.ocean} where
the case of a satellite with a subsurface ocean is treated. Here, the two models of the same problem
are used to estimate the error made by the approximation.

\subsection{Poincar\'e-Hough model}
\label{sec.poincare-hough}
As in the previous model, $A$, $B$, $C$ designate the principal moments of inertia of the whole
satellite. Those of the liquid core are denote by $A_c$, $B_c$, $C_c$.  We assume that the axes of
the core/mantle ellipsoidal boundary are aligned to those of the satellite surface.  Hence, the
principal axes $(\vec I_c, \vec J_c, \vec K_c)$ of the core are aligned to those of the mantle
denoted $(\vec I_m, \vec J_m, \vec K_m)$ which are also aligned to those of the whole satellite
$(\vec I, \vec J, \vec K)$. The vector $\vec \omega$ still represents the rotation vector of $(\vec
I, \vec J, \vec K)$ with respect to the laboratory frame expressed in the laboratory frame. We add
the rotation vector $\vec \omega'_c$ associated with the {\em simple motion} of the liquid
core with respect to the mantle and expressed in the mantle-fixed frame \citep{Poincare10}. As in
the rigid case, the laboratory frame rotates with respect to the inertial frame at the speed $\vec
\Omega$. Let $\inertia$, $\inertia'_c$ and $\inertia'$ be the inertia matrices defined as
\begin{eqnarray}
\label{eq.In}
&& \inertia = \mat R\diag(A, B, C)\trans{\mat R} , \\
\label{eq.Inc}
&& \inertia'_c = \diag(A_c, B_c, C_c) , \\
\label{eq.In'}
&& \inertia' = \diag(A', B', C')\trans{\mat R} ,
\end{eqnarray}
where $\mat R = [\vec I, \vec J, \vec K]$ is the rotation matrix of the mantle relative to the
laboratory frame. Furthermore, we have defined
\begin{equation}
A' = A_c\sqrt{1-\alpha_c^2} , \qquad
B' = B_c\sqrt{1-\beta_c^2} , \qquad
C' = C_c\sqrt{1-\gamma_c^2} ,
\label{eq.ABC'}
\end{equation}
with
\begin{equation}
\alpha_c = \frac{C_c-B_c}{A_c} ,\qquad
\beta_c = \frac{C_c-A_c}{B_c} ,\qquad
\gamma_c = \frac{B_c-A_c}{C_c} .
\end{equation}
For this problem, the Lie velocity is $\vec \eta = (\vec \omega, \vec \omega'_c)$ and the
generalised coordinates are limited to $\vec q = (\vec I, \vec J, \vec K)$. Coordinates
associated with the simple motion of the liquid core do not appear in the equations of
motion because the fluid is assumed to be incompressible and its volume is set by the mantle, thus
the kinetic and the potential energies only depends on $\vec \eta$ and $\vec q$.  The kinetic energy
$T_\mathrm{fc}(\vec \eta, \vec q)$ of rotation of the satellite is \citep{Poincare10, Hough95}
\begin{equation}
T_\mathrm{fc}(\vec \eta, \vec q) 
 =  \frac{\trans{(\vec \omega+\vec \Omega)} \inertia (\vec \omega+\vec \Omega)}{2}
  + \frac{\trans{\vec \omega'}_c \inertia'_c \vec \omega'_c}{2}
  + \trans{\vec \omega'}_c \inertia' (\vec \omega+\vec \Omega) ,
\label{eq.Tfc}
\end{equation}
The potential energy is the same as in the rigid satellite case (see
sect.~\ref{sec.rigid-satellite}). Thus, the Lagrangian $L_\mathrm{fc}(\vec \eta, \vec q)$ reads
\begin{equation}
L_\mathrm{fc}(\vec \eta, \vec q) 
  = \frac{\trans{(\vec \omega+\vec \Omega)} \inertia (\vec \omega+\vec \Omega)}{2}
  + \frac{\trans{\vec \omega'}_c \inertia'_c \vec \omega'_c}{2}
  + \trans{\vec \omega'}_c \inertia'(\vec \omega + \vec \Omega)
  - \frac{3{\cal G}M_p}{2}\frac{\trans{\vec r}\inertia\vec r}{r^5} .
\end{equation}
The Lie momenta associated with $\vec \omega$ and $\vec \omega'_c$ are respectively
\begin{eqnarray}
&&\vec \Pi = \frac{\partial L_\mathrm{fc}}{\partial \vec \omega} = 
\inertia
(\vec \omega+\vec \Omega) + \trans{\inertia'} \vec \omega'_c ,
\\
&&\vec \Pi'_c = \frac{\partial L_\mathrm{fc}}{\partial \vec \omega'_c} = \inertia'_c \vec
\omega'_c + \inertia'(\vec \omega + \vec \Omega) ,
\end{eqnarray}
with the inverse transformation,
\begin{eqnarray}
&&\vec \omega = \invinertia\vec \Pi-\trans{\invinertia'}\vec \Pi'_c - \vec \Omega , \\
&&\vec \omega'_c = \invinertia'_c\vec \Pi'_c - \invinertia'\vec \Pi ,
\end{eqnarray}
where
\begin{eqnarray}
&& \invinertia
   = \mat R \diag\left(\frac{A_c}{AA_c-A'^2}, \frac{B_c}{BB_c-B'^2}, \frac{C_c}{CC_c-C'^2}\right) \trans{\mat R}, \\
&& \invinertia'_c
   = \diag\left(\frac{A  }{AA_c-A'^2}, \frac{B  }{BB_c-B'^2}, \frac{C  }{CC_c-C'^2}\right) , \\
&& \invinertia'
   = \diag\left(\frac{A' }{AA_c-A'^2}, \frac{B' }{BB_c-B'^2}, \frac{C' }{CC_c-C'^2}\right) \trans{\mat R} .
\end{eqnarray}
The Hamiltonian of the problem is then
\begin{equation}
H_\mathrm{fc}(\vec y, t) = \frac{\trans{\vec \Pi}\invinertia \vec \Pi}{2}
           +\frac{\trans{\vec \Pi'}_c\invinertia'_c \vec \Pi'_c}{2}
           -      \trans{\vec \Pi'}_c\invinertia'\vec \Pi
           - \trans{\vec \Omega}\vec \Pi
           +\frac{3{\cal G}M_p}{2}\frac{\trans{\vec r} \inertia \vec r}{r^5} ,
\end{equation}
with the state vector $\vec y = (\vec \Pi'_c, \vec \Pi, \vec I, \vec J, \vec K)$.  In these
variables, the Poisson matrix reads
\begin{equation}
\mat B_\mathrm{fc}(\vec y) = \begin{bmatrix}
\hvec \Pi'_c & \mat 0 & \mat 0 & \mat 0 & \mat 0 \\
\mat 0 & \hvec \Pi & \hvec I & \hvec J & \hvec K \\
\mat 0 & \hvec I & \mat 0 & \mat 0 & \mat 0 \\
\mat 0 & \hvec J & \mat 0 & \mat 0 & \mat 0 \\
\mat 0 & \hvec K & \mat 0 & \mat 0 & \mat 0
\end{bmatrix}
\end{equation}
and the equations of motion are
\begin{eqnarray}
&&\dot{\vec \Pi'}_c = \vec \omega'_c \times \vec \Pi'_c , \\
&&\dot{\vec \Pi} = \vec \Pi \times \vec \Omega - 3 \frac{{\cal G}M_p}{r^5}(\inertia \vec r)\times \vec r , \\
&&\dot{\vec I} = \vec \omega \times \vec I , \\
&&\dot{\vec J} = \vec \omega \times \vec J , \\
&&\dot{\vec K} = \vec \omega \times \vec K .
\end{eqnarray}
As in the rigid case (Sect.~\ref{sec.rigid-satellite}), we now split the Hamiltonian
$H_\mathrm{fc}(\vec y, t)$ into its autonomous part $H_\mathrm{fc}^0(\vec y)$ and a perturbation
$H_\mathrm{fc}^1(\vec y, t)$ using the decomposition of the gravitational potential energy $U_0(\vec
y)$ and $U_1(\vec y, t)$, Eqs.~(\ref{eq.U0}-\ref{eq.U1}).  There are seven Casimir functions given
by
\begin{equation}
\begin{split}
&
C_\mathrm{fc}^0(\vec y) = \frac{1}{2}\vec \Pi'_c \cdot \vec \Pi'_c, 
\\ &
C_\mathrm{fc}^1(\vec y) = \frac{1}{2}\vec I\cdot \vec I,
\quad
C_\mathrm{fc}^2(\vec y) = \frac{1}{2}\vec J\cdot \vec J,
\quad
C_\mathrm{fc}^3(\vec y) = \frac{1}{2}\vec K\cdot \vec K,
\\ &
C_\mathrm{fc}^4(\vec y) = \vec J \cdot \vec K,
\quad
C_\mathrm{fc}^5(\vec y) = \vec K \cdot \vec I,
\quad
C_\mathrm{fc}^6(\vec y) = \vec I \cdot \vec J.
\end{split}
\end{equation}
The equilibrium $\vec y_e$ of $H_\mathrm{fc}^0(\vec y)$ is solution of
\begin{eqnarray}
&&\vec \omega'_{c,e} - \mu_0 \vec \Pi'_{c,e} = \vec 0 , \\
&&\vec \omega_e = \vec 0 , \\
\label{eq.nablaFI}
&&\frac{A_c(\vec I_e\cdot \vec \Pi_e) - A'(\vec I_e\cdot \vec \Pi'_{c,e})}{AA_c-A'^2}\vec \Pi_e + 3A\mat
S_0 \vec I_e - \mu_1 \vec I_e - \mu_5\vec K_e - \mu_6 \vec J_e = \vec 0 , \\
\label{eq.nablaFJ}
&&\frac{B_c(\vec J_e\cdot \vec \Pi_e) - B'(\vec J_e\cdot \vec \Pi'_{c,e})}{BB_c-B'^2}\vec \Pi_e + 3B\mat
S_0 \vec J_e - \mu_2 \vec J_e - \mu_4\vec K_e - \mu_6 \vec I_e = \vec 0 , \\
\label{eq.nablaFK}
&&\frac{C_c(\vec K_e\cdot \vec \Pi_e) - C'(\vec K_e\cdot \vec \Pi'_{c,e})}{CC_c-C'^2}\vec \Pi_e + 3C\mat
S_0 \vec K_e - \mu_3 \vec K_e - \mu_4\vec J_e - \mu_5 \vec I_e = \vec 0 . \qquad
\end{eqnarray}
We stress that $\vec \Pi$ is written in the laboratory frame while $\vec \Pi'_c$ is expressed in the
mantle-fixed frame. Thus, in Eq.~(\ref{eq.nablaFI}), $(\vec I\cdot\vec \Pi) = I_x\Pi_x + I_y\Pi_y +
I_z\Pi_z$ whereas $(\vec I \cdot \vec \Pi'_c) = \Pi'_{c,x}$. The same reasoning holds in
Eqs.~(\ref{eq.nablaFJ},\ref{eq.nablaFK}).  The norm of the angular velocity $\vec \omega'_{c,e}$ can
be arbitrarily chosen.  This is due to the conservation of the Casimir $C_\mathrm{fc}^0(\vec y)$.
Here, we assume that the fluid core has no mean angular velocity with respect to the mantle and thus
$\vec \omega'_{c,e} = \vec 0$. Under this hypothesis, we get
\begin{equation}
\begin{split}
\vec \omega_e = \vec 0, \quad
\vec \Pi'_{c,e} = C'\Omega \vec k, \quad
\vec \Pi_e = C\Omega \vec k, \quad
\vec I_e = \vec i, \quad
\vec J_e = \vec j, \quad
\vec K_e = \vec k, \\
\mu_0 = 0, \quad
\mu_1 = 3A\sigma^0_{xx}, \quad
\mu_2 = 3B\sigma^0_{yy}, \quad
\mu_3 = 3C\sigma^0_{zz} + C\Omega^2,
\mu_4 = \mu_5 = \mu_6 = 0\ .
\end{split}
\end{equation}
The linear system is expressed in the coordinates
\begin{equation}
\begin{split}
\delta\vec y^* = \Big(\delta\Pi_{z}, \delta I_{y}, \delta \Pi'_{c,x},
\delta \Pi'_{c,y}, \delta \Pi_{x}, \delta \Pi_{y}, \delta I_{z},
\delta J_{z},
\delta \Pi'_{c,z}, \delta I_{x}, 
\delta J_{y}, \delta K_{z}, \\
\delta I_{y}+\delta J_{x}, \delta I_{z}+\delta K_{x},
\delta J_{z} + \delta K_{y}\trans{\Big)} .
\end{split}
\label{eq.deltaybis}
\end{equation}
Let $\mat A_\mathrm{fc}^*(\vec y_e)$ be the matrix of the linear system evaluated at the equilibrium
point and expressed in the coordinates $\delta \vec y^*$. As in the rigid case, we define the
matrices $\mat A_\mathrm{fc}^1$ and $\mat A_\mathrm{fc}^2$ such that
\begin{equation}
\mat A_\mathrm{fc}^*(\vec y_e) = \begin{bmatrix}
\mat A_\mathrm{fc}^1 & \mat 0 & \cdot \\
\mat 0 & \mat A_\mathrm{fc}^2 & \cdot \\
\mat 0 & \mat 0 & \mat 0 
\end{bmatrix} ,
\label{eq.matAfc}
\end{equation}
where the dots $\cdot$ still denote arbitrary matrices. We get
\begin{equation}
\mat A_\mathrm{fc}^1 = \begin{bmatrix}
0 & -3(B-A)(\sigma^0_{xx}-\sigma^0_{yy}) \\
\Frac{C_c}{CC_c-C'^2} & 0 
\end{bmatrix} ,
\label{eq.A1bis}
\end{equation}
and
\begin{equation}
\mat A_\mathrm{fc}^2 = 
\begin{bmatrix}
0 & \Frac{C'}{\reduced B}\Omega & 0 & -\Frac{C'}{\reduced B'}\Omega & 0 &
-\Frac{C'}{\reduced B'}C\Omega^2 \\[0.8em]
-\Frac{C'}{\reduced A}\Omega & 0 & \Frac{C'}{\reduced A'}\Omega & 0 &
\Frac{C'}{\reduced A'}C\Omega^2 & 0 \\[0.8em]
0 & 0 & 0 & \Omega & 0 & -3(C-B)\kappa_2 \\[0.8em]
0 & 0 & -\Omega & 0 & 3(C-A)\kappa_1 & 0 \\[0.8em]
0 & \Frac{1}{\reduced B'} & 0 & -\Frac{1}{\reduced B_c} & 0 &
\left(1-\Frac{C}{\reduced B_c}\right)\Omega \\[0.8em]
-\Frac{1}{\reduced A'} & 0 & \Frac{1}{\reduced A_c} & 0 &
-\left(1-\Frac{C}{\reduced A_c}\right)\Omega & 0
\end{bmatrix}
\label{eq.A2bis}
\end{equation}
with
\begin{equation}
\begin{split}
&
\Frac{1}{\reduced A  } = \frac{A  }{AA_c-A'^2}, \quad
\Frac{1}{\reduced A_c} = \frac{A_c}{AA_c-A'^2}, \quad
\Frac{1}{\reduced A' } = \frac{A' }{AA_c-A'^2}, \\
&
\Frac{1}{\reduced B  } = \frac{B  }{BB_c-B'^2}, \quad
\Frac{1}{\reduced B_c} = \frac{B_c}{BB_c-B'^2}, \quad
\Frac{1}{\reduced B' } = \frac{B' }{BB_c-B'^2}, \\
&
\Frac{1}{\reduced C  } = \frac{C  }{CC_c-C'^2}, \quad
\Frac{1}{\reduced C_c} = \frac{C_c}{CC_c-C'^2}, \quad
\Frac{1}{\reduced C' } = \frac{C' }{CC_c-C'^2}.
\end{split}
\end{equation}
The eigenfrequencies are
\begin{eqnarray}
\label{eq.lofc}
&& \omega_{\mathrm{fc},u} = \left(\frac{CC_c}{CC_c-C'^2}\right)^{1/2}
\omega_{\mathrm{rs},u} , \\
\label{eq.wofc}
&& \omega_{\mathrm{fc},v} = \omega_{\mathrm{rs},v} +
O(\epsilon) , \\
\label{eq.lafc}
&& \omega_{\mathrm{fc},w} = \omega_{\mathrm{rs},w} +
O(\epsilon) , \\
\label{eq.cofc}
&& \omega_{\mathrm{fc},z} = \frac{C'}{\sqrt{A_cB_c}}\Omega +
O(\epsilon)
\end{eqnarray}
with $\epsilon$ being the mass of the core divided by the total mass of the satellite.
$\omega_{\mathrm{rs},u}$, $\omega_{\mathrm{rs},v}$, and $\omega_{\mathrm{rs},w}$ are the frequencies
obtained in the rigid case (Eqs.~\ref{eq.nulong},\ref{eq.nuwl}). $\omega_{\mathrm{fc},z}$ is the
frequency of the additional degree of freedom induced by the presence of the liquid core. In the
case where the fluid core represents a significant fraction of the total mass of the satellite,
Eqs.~(\ref{eq.wofc}-\ref{eq.cofc}) are no longer valid and eigenfrequencies should be directly
computed from the matrix $\mat A_\mathrm{fc}^2$ (Eq.~\ref{eq.A2bis}).

The Lyapunov function (Eq.~\ref{eq.Lyap}) associated with this problem is
\begin{equation}
\begin{split}
N_\mathrm{fc}(\vec y) = &
 \frac{1}{2\reduced A_c} \left(\Pi_x-\frac{A'}{A_c}\Pi'_{c,x}+\frac{1}{2}\left(C-\reduced A_c\right)(I_z-K_x)\right)^2
\\ &
 + \frac{1}{2\reduced B_c} \left(\Pi_y-\frac{B'}{B_c}\Pi'_{c,y}+\frac{1}{2}\left(C-\reduced B_c\right)(J_z-K_y)\right)^2
 + \frac{1}{2\reduced C_c} \Pi_z^2
\\ &
 + \frac{1}{2A_c} \left(\Pi'_{c,x}-\frac{1}{2}A'\Omega(I_z-K_x)\right)^2
 + \frac{1}{2B_c} \left(\Pi'_{c,y}-\frac{1}{2}B'\Omega(J_z-K_y)\right)^2
\\ &
 + \frac{1}{2}n_1(I_y-J_x)^2
 + \frac{1}{2}n_2(I_z-K_x)^2
 + \frac{1}{2}n_3(J_z-K_y)^2 ,
\end{split}
\end{equation}
where $n_1$, $n_2$, and $n_3$ are the same as in the rigid case (see Eq.~\ref{eq.n123}). Given that
$\reduced A_c$, $\reduced B_c$, $\reduced C_c$, $A_c$, and $B_c$ are all positive, the nonlinear
stability criterion is identical to that of a rigid satellite, namely $C>B>A$. In particular, there
is no restriction on the moments of inertia of the core ($A_c, B_c, C_c$).

The driven equations of motion of the satellite with a liquid core in the vicinity of the relative
equilibrium $\vec y_e$ are of the form
\begin{equation}
\delta \dot{\vec y}^k - \mat A_\mathrm{fc}^k \delta \vec y^k = \delta
\vec z_\mathrm{fc}^k(t) , \quad k=1,2 ,
\end{equation}
with
\begin{equation}
\delta \vec y^1 = \trans{(\delta \Pi_z, \delta I_y)} , \qquad
\delta \vec y^2 = \trans{(\delta \Pi'_{c,x}, \delta \Pi'_{c,y}, \delta
\Pi_x, \delta \Pi_y, \delta I_z, \delta J_z)} ,
\end{equation}
and
\begin{equation}
\delta \vec z_\mathrm{fc}^1(t) = \begin{pmatrix}
3(B-A)\sigma^1_{xy}(t) \\ 0
\end{pmatrix} , \qquad
\delta \vec z_\mathrm{fc}^2(t) = \begin{pmatrix}
3(C-B)\sigma^1_{yz}(t) \\ -3(C-A)\sigma^1_{xz}(t) \\ 0 \\ 0 \\ 0 \\ 0
\end{pmatrix} .
\label{eq.deltazbis}
\end{equation}

\subsection{Quasi-spherical approximation}
\label{sec.fluid-approximation}
In this section, we reconsider the case of a satellite with a liquid core, but we assimilate $A'$,
$B'$ and $C'$ to the moments of inertia of the core, i.e., we assume
\begin{equation}
A'\approx A_c , \qquad
B'\approx B_c , \qquad
C'\approx C_c .
\label{eq.approx}
\end{equation}
According to Eq.~(\ref{eq.ABC'}), this is equivalent to a first order approximation in $\alpha_c$,
$\beta_c$ and $\gamma_c$. With this simplification, the kinetic energy (Eq.~\ref{eq.Tfc}) can be
rewritten as follows
\begin{equation}
T_\mathrm{fc'}(\vec \eta, \vec q) = 
  \frac{\trans{(\vec \omega + \vec \Omega)} \inertia_m (\vec \omega + \vec
\Omega)}{2} + \frac{\trans{(\vec \omega'_c + \trans{\mat R}(\vec \omega
+ \vec \Omega))}\inertia'_c (\vec \omega'_c + \trans{\mat R}(\vec \omega
+ \vec \Omega))}{2}
\label{eq.Tfc'}
\end{equation}
where
\begin{equation}
\begin{split}
\inertia_m &= \inertia - \mat R \inertia'_c \trans{\mat R}  \\
           &= \mat R \diag(A_m, B_m, C_m) \trans{\mat R}
\end{split}
\end{equation}
is the inertia tensor of the mantle written in the laboratory frame ($A_m=A-A_c$, $B_m=B-B_c$, and
$C_m=C-C_c$). According to the expression (\ref{eq.Tfc'}), the problem behaves as if the liquid core
were rotating rigidly relative to the mantle at the angular velocity $\vec \omega'_c$ with a matrix
of inertia $\inertia'_c$ constant in the mantle-fixed frame.  Indeed, $\vec \omega'_c + \trans{\mat
R}(\vec \omega + \vec \Omega)$ is the rotation speed of the core with respect to the inertial frame
written in the mantle-fixed frame. We here retrieve the approximation made by
\citet{Mathews91} who neglected the small departure of the fluid velocity field from a pure solid
rotation. Following the same procedure as in Sect.~\ref{sec.poincare-hough}, the two submatrices of
the linearised system written in the set of variables $\delta{\vec y}^*$ (Eq.~\ref{eq.deltaybis})
become
\begin{equation}
\mat A_\mathrm{fc'}^1 = \begin{bmatrix}
0 & -3(B-A)(\sigma^0_{xx}-\sigma^0_{yy}) \\
\Frac{1}{C_m} & 0 
\end{bmatrix} ,
\label{eq.A1ter}
\end{equation}
and
\begin{equation}
\mat A_\mathrm{fc'}^2 = \begin{bmatrix}
0 & \Frac{BC_c}{B_mB_c}\Omega & 0 & -\Frac{C_c}{B_m}\Omega & 0 &
-\Frac{CC_c}{B_m}\Omega^2 \\[1.05em]
-\Frac{AC_c}{A_mA_c}\Omega & 0 & -\Frac{C_c}{A_m}\Omega & 0 & \Frac{CC_c}{A_m}\Omega^2 & 0 \\[1.05em]
0 & 0 & 0 & \Omega & 0 & -3(C-B)\kappa_2 \\
0 & 0 & -\Omega & 0 & 3(C-A)\kappa_1 & 0 \\
0 & \Frac{1}{B_m} & 0 & -\Frac{1}{B_m} & 0 & \left(1-\Frac{C}{B_m}\right)\Omega \\
-\Frac{1}{A_m} & 0 & \Frac{1}{A_m} & 0 & -\left(1-\Frac{C}{A_m}\right)\Omega & 0
\end{bmatrix} .
\label{eq.A2ter}
\end{equation}
Although we retrieve the eigenfrequencies obtained in section~\ref{sec.poincare-hough} within the
approximation (Eq.~\ref{eq.approx}) only, the second member $\delta{\vec z}_\mathrm{fc'}(t)$ of the
driven system is exactly the same as $\delta{\vec z}_\mathrm{fc}(t)$ (Eq.~\ref{eq.deltazbis}).

\section{Satellite with a subsurface ocean}
\label{sec.ocean}
Here, we consider a satellite with a rigid central part $c$ (also called interior) and a rigid shell
$s$ separated by a global ocean $o$. By assumption, the shell is ellipsoidal with inner radii $a_o,
b_o, c_o$ and outer radii $a_s, b_s, c_s$. The interior, an ellipsoid of radii $a_c, b_c, c_c$,
might be differentiated, i.e., it can be made of a succession of $N$ concentric ellipsoidal layers
with different densities $(\rho_{i})_{1\leq i \leq N}$ and outer radii $a_{i}, b_{i}, c_{i}$. We
have thus $a_{N} = a_c$, $b_{N} = b_c$ and $c_{N} = c_c$.  The ocean and the shell are assumed to be
homogeneous with respective density $\rho_o$ and $\rho_s$. Nevertheless, the results can easily be
extended to the case of a stratified rigid shell. Because the simple motion introduced by
\citet{Poincare10} for a satellite with a liquid core cannot be applied in this case, we use the
approximation described in Sect.~\ref{sec.fluid-approximation}. We could describe the evolution of
the central region and of the ocean in the shell-fixed frame to remain close to the study made on
the satellite with a liquid core, but equations are more symmetrical if all coordinates are given
with respect to a same given frame which we chose to be the laboratory frame.  In this frame, the
configuration of the system is given by the coordinates of the principal axes of the interior and
the shell, i.e., the generalised coordinates are $\vec q = (\vec I_c, \vec J_c, \vec K_c, \vec I_s,
\vec J_s, \vec K_s)$.  The Lie velocities are the rotation vectors of the three layers with respect
to the laboratory frame $\vec \eta = (\vec \omega_o, \vec \omega_c, \vec \omega_s)$. Within the
approximation of Sect.~\ref{sec.fluid-approximation}, the kinetic energy of the satellite with a
global ocean reads
\begin{equation}
\begin{split}
T_\mathrm{go}(\vec \eta, \vec q)
  = \frac{\trans{(\vec \omega_c+\vec \Omega)}\inertia_c (\vec \omega_c + \vec \Omega)}{2}
  + \frac{\trans{(\vec \omega_s+\vec \Omega)}\inertia_s (\vec \omega_s + \vec \Omega)}{2} \\
  + \frac{\trans{(\vec \omega_o+\vec \Omega)}\inertia_o (\vec \omega_o+\vec \Omega)}{2} ,
\end{split}
\end{equation}
with the inertia tensors
\begin{eqnarray}
&&\inertia_c = \mat R_c \diag(A_c, B_c, C_c) \trans{\mat R}_c , \\
&&\inertia_s = \mat R_s \diag(A_s, B_s, C_s) \trans{\mat R}_s , \\
&&\inertia_o = \mat R_s \diag(A'_s, B'_s, C'_s) \trans{\mat R}_s
             - \mat R_c \diag(A'_c, B'_c, C'_c) \trans{\mat R}_c ,
\end{eqnarray}
where $\mat R_c = [\vec I_c, \vec J_c, \vec K_c]$, $\mat R_s = [\vec I_s,
\vec J_s, \vec K_s]$, and
\begin{eqnarray}
\label{eq.Ac}
&&A_c = \sum_{i=1}^N\frac{4\pi}{15}\rho_i\left(a_ib_ic_i(b^2_i+c^2_i)-a_{i-1}b_{i-1}c_{i-1}(b^2_{i-1}+c^2_{i-1})\right) , \\
\label{eq.Am}
&&A_s = \frac{4\pi}{15}\rho_s\left(a_sb_sc_s(b^2_s+c^2_s)-a_ob_oc_o(b_o^2+c_o^2)\right) ,\\
\label{eq.A'c}
&&A'_c = \frac{4\pi}{15}\rho_oa_cb_cc_c(b_c^2+c_c^2) , \\
\label{eq.A'm}
&&A'_s = \frac{4\pi}{15}\rho_oa_ob_oc_o(b_o^2+c_o^2) .
\end{eqnarray}
In Eq.~(\ref{eq.Ac}), we apply the convention 
$a_0 = b_0 = c_0 = 0$.
The other quantities $B$, $C$ are deduced from Eqs.~(\ref{eq.Ac}-\ref{eq.A'm}) by circular
permutation of $a$, $b$, $c$. Let us stress that the matrix of inertia of the whole satellite is
simply
\begin{equation}
\inertia = \inertia_c + \inertia_s + \inertia_o .
\end{equation}

In addition to the gravitational potential energy $U(\vec y, t)$ between the planet point mass and
the extended satellite, to get the Lagrangian we also need to include the self gravitational
potential energy $U_\mathrm{self}(\vec q)$ of the satellite as it is a function of the relative
orientation of the interior and the shell. This potential energy reads \citep{Laplace98}
\begin{equation}
\begin{split}
U_\mathrm{self}(\vec q) 
                &= \frac{u_{xx}}{2}(\vec I_c \cdot \vec I_s)^2
                 + \frac{u_{xy}}{2}(\vec I_c \cdot \vec J_s)^2
                 + \frac{u_{xz}}{2}(\vec I_c \cdot \vec K_s)^2 \\
                &+ \frac{u_{yx}}{2}(\vec J_c \cdot \vec I_s)^2
                 + \frac{u_{yy}}{2}(\vec J_c \cdot \vec J_s)^2
                 + \frac{u_{yz}}{2}(\vec J_c \cdot \vec K_s)^2 \\
                &+ \frac{u_{zx}}{2}(\vec K_c \cdot \vec I_s)^2
                 + \frac{u_{zy}}{2}(\vec K_c \cdot \vec J_s)^2
                 + \frac{u_{zz}}{2}(\vec K_c \cdot \vec K_s)^2 ,
\end{split}
\end{equation}
with 
\begin{eqnarray}
&
u_{xx} = \frac{8\pi}{15}{\cal G}\left(\rho_s f_s + (\rho_o-\rho_s)f_o\right) \sum_{i=1}^N (\rho_i-\rho_{i+1})a_i^3b_ic_i , 
\label{eq.uxyza}
\\[0.2em] &
u_{xy} = \frac{8\pi}{15}{\cal G}\left(\rho_s g_s + (\rho_o-\rho_s)g_o\right) \sum_{i=1}^N (\rho_i-\rho_{i+1})a_i^3b_ic_i ,
\\[0.2em] &
u_{xz} = \frac{8\pi}{15}{\cal G}\left(\rho_s h_s + (\rho_o-\rho_s)h_o\right) \sum_{i=1}^N (\rho_i-\rho_{i+1})a_i^3b_ic_i ,
\\[0.2em] &
u_{yx} = \frac{8\pi}{15}{\cal G}\left(\rho_s f_s + (\rho_o-\rho_s)f_o\right) \sum_{i=1}^N (\rho_i-\rho_{i+1})a_ib_i^3c_i ,
\\[0.2em] &
u_{yy} = \frac{8\pi}{15}{\cal G}\left(\rho_s g_s + (\rho_o-\rho_s)g_o\right) \sum_{i=1}^N (\rho_i-\rho_{i+1})a_ib_i^3c_i ,
\\[0.2em] &
u_{yz} = \frac{8\pi}{15}{\cal G}\left(\rho_s h_s + (\rho_o-\rho_s)h_o\right) \sum_{i=1}^N (\rho_i-\rho_{i+1})a_ib_i^3c_i ,
\\[0.2em] &
u_{zx} = \frac{8\pi}{15}{\cal G}\left(\rho_s f_s + (\rho_o-\rho_s)f_o\right) \sum_{i=1}^N (\rho_i-\rho_{i+1})a_ib_ic_i^3 ,
\\[0.2em] &
u_{zy} = \frac{8\pi}{15}{\cal G}\left(\rho_s g_s + (\rho_o-\rho_s)g_o\right) \sum_{i=1}^N (\rho_i-\rho_{i+1})a_ib_ic_i^3 ,
\\[0.2em] &
u_{zz} = \frac{8\pi}{15}{\cal G}\left(\rho_s h_s + (\rho_o-\rho_s)h_o\right) \sum_{i=1}^N (\rho_i-\rho_{i+1})a_ib_ic_i^3 ,
\label{eq.uxyzb}
\end{eqnarray}
where $\rho_{N+1} \equiv \rho_o$ and for $*\in\{s,o\}$, 
\begin{eqnarray}
&&f_* = 2\pi \frac{a_*b_*}{c_*^2} \int_0^1 
{\left(1+\frac{a_*^2-c_*^2}{c_*^2}t^2\right)^{-3/2}\left(1+\frac{b_*^2-c_*^2}{c_*^2}t^2\right)^{-1/2}}{t^2\,\dd t} , \\
&&g_* = 2\pi \frac{a_*b_*}{c_*^2} \int_0^1 
{\left(1+\frac{a_*^2-c_*^2}{c_*^2}t^2\right)^{-1/2}\left(1+\frac{b_*^2-c_*^2}{c_*^2}t^2\right)^{-3/2}}{t^2\,\dd t} , \\
&&h_* = 2\pi \frac{a_*b_*}{c_*^2} \int_0^1 
{\left(1+\frac{a_*^2-c_*^2}{c_*^2}t^2\right)^{-1/2}\left(1+\frac{b_*^2-c_*^2}{c_*^2}t^2\right)^{-1/2}}{t^2\,\dd t} .
\end{eqnarray}
The Lagrangian $L_\mathrm{go}(\vec \eta, \vec q)$ of the problem is then
\begin{equation}
\begin{split}
L_\mathrm{go}(\vec \eta, \vec q) = 
    \frac{\trans{(\vec \omega_c+\vec \Omega)}\inertia_c (\vec \omega_c + \vec \Omega)}{2}
  + \frac{\trans{(\vec \omega_s+\vec \Omega)}\inertia_s (\vec \omega_s + \vec \Omega)}{2} \\
  + \frac{\trans{(\vec \omega_o+\vec \Omega)}\inertia_o (\vec \omega_o+\vec \Omega)}{2} 
  - \frac{3{\cal G}M_p}{2}\frac{\trans{\vec r} \inertia \vec r}{r^5} -
    U_\mathrm{self}(\vec q) .
\end{split}
\end{equation}
The Lie momenta associated with $\vec \eta = (\vec \omega_o, \vec \omega_c, \vec \omega_s)$
are
\begin{eqnarray}
&& \vec \Pi_o = \Frac{\partial L_\mathrm{go}}{\partial \vec \omega_o} = \inertia_o(\vec \omega_o + \vec \Omega) , \\
&& \vec \Pi_c = \Frac{\partial L_\mathrm{go}}{\partial \vec \omega_c} = \inertia_c(\vec \omega_c + \vec \Omega) , \\
&& \vec \Pi_s = \Frac{\partial L_\mathrm{go}}{\partial \vec \omega_s} = \inertia_s(\vec \omega_s + \vec \Omega) ,
\end{eqnarray}
from which we deduce the Hamiltonian
\begin{equation}
\begin{split}
H_\mathrm{go}(\vec y) 
  = \frac{\trans{\vec \Pi}_c (\inertia_c)^{-1}\vec \Pi_c}{2}
  + \frac{\trans{\vec \Pi}_o (\inertia_o)^{-1}\vec \Pi_o}{2}
  + \frac{\trans{\vec \Pi}_s (\inertia_s)^{-1}\vec \Pi_s}{2} \\
  - \trans{\vec \Omega}(\vec \Pi_c + \vec \Pi_o + \vec \Pi_s)
+ \frac{3{\cal G}M_p}{2}\frac{\trans{\vec r}\inertia \vec r}{r^5} +
U_\mathrm{self}(\vec q) ,
\end{split}
\label{eq.Hgo}
\end{equation}
which is a function of $\vec y = (\vec \Pi_o, \vec y_c, \vec y_s)$ with $\vec y_i = (\vec \Pi_i,
\vec I_i, \vec J_i, \vec K_i)$. The Poisson matrix $\mat B_\mathrm{go}(\vec y)$ associated
with this set of variables is
\begin{equation}
\mat B_\mathrm{go}(\vec y) = \begin{bmatrix}
\hvec \Pi_o & \mat 0 & \mat 0 \\
\mat 0 & \mat b(\vec y_c) & \mat 0 \\
\mat 0 & \mat 0 & \mat b(\vec y_s)
\end{bmatrix} ,\qquad
\mat b(\vec y_i) = \begin{bmatrix}
\hvec \Pi_i & \hvec I_i & \hvec J_i & \hvec K_i \\
\hvec I_i & \mat 0 & \mat 0 & \mat 0 \\
\hvec J_i & \mat 0 & \mat 0 & \mat 0 \\
\hvec K_i & \mat 0 & \mat 0 & \mat 0
\end{bmatrix} , \qquad
i=c,s .
\end{equation}
Although $\vec y$ has 27 components, the system evolves in a phase space $\Sigma_\mathrm{go}$ of
dimension $14=2\times7$ whose degrees of freedom are the three rotations of the central region, the
three rotation of the shell and an additional degree of freedom associated with the ocean:
\begin{equation}
\Sigma_\mathrm{go} = \{\vec y \in \mathbb{R}^{27} \colon C_\mathrm{go}^i(\vec y) = c_i, 0\leq i
\leq 12 \}
\end{equation}
where the thirteen Casimir functions are
\begin{equation}
\begin{array}{llll}
C_\mathrm{go}^{ 0}(\vec y) = \Frac{1}{2}\trans{\vec \Pi}_o \vec \Pi_o , \ &
C_\mathrm{go}^{ 1}(\vec y) = \Frac{1}{2}\trans{\vec I}_c \vec I_c , \ &
C_\mathrm{go}^{ 2}(\vec y) = \Frac{1}{2}\trans{\vec J}_c \vec J_c , \ &
C_\mathrm{go}^{ 3}(\vec y) = \Frac{1}{2}\trans{\vec K}_c \vec K_c , \\[0.5em]
C_\mathrm{go}^{ 4}(\vec y) =            \trans{\vec J}_c \vec K_c , \ &
C_\mathrm{go}^{ 5}(\vec y) =            \trans{\vec K}_c \vec I_c , \ &
C_\mathrm{go}^{ 6}(\vec y) =            \trans{\vec I}_c \vec J_c , \ &
C_\mathrm{go}^{ 7}(\vec y) = \Frac{1}{2}\trans{\vec I}_s \vec I_s , \\[0.5em]
C_\mathrm{go}^{ 8}(\vec y) = \Frac{1}{2}\trans{\vec J}_s \vec J_s , \ &
C_\mathrm{go}^{ 9}(\vec y) = \Frac{1}{2}\trans{\vec K}_s \vec K_s , \ &
C_\mathrm{go}^{10}(\vec y) =            \trans{\vec J}_s \vec K_s , \ &
C_\mathrm{go}^{11}(\vec y) =            \trans{\vec K}_s \vec I_s , \\[0.8em]
C_\mathrm{go}^{12}(\vec y) =            \trans{\vec I}_s \vec J_s .
\end{array}
\end{equation}
In order to proceed, we have to compute the inverse of the inertia matrix of the ocean
$(\inertia_o)^{-1}$ for which we are missing the principal basis. The other terms of the Hamiltonian
$H_\mathrm{go}$ (Eq.~\ref{eq.Hgo}) are fully explicit and do not cause any problem.  To make the
computation analytical, we anticipate the equilibrium point solution
\begin{equation}
\begin{split}
&\vec \Pi_{o,e} = C_o\Omega \vec k \ \quad
\vec \Pi_{s,e} = C_s\Omega \vec k , \quad
  \vec I_{s,e} = \vec i , \quad
  \vec J_{s,e} = \vec j , \quad
  \vec K_{s,e} = \vec k , \\
&\vec \Pi_{c,e} = C_c\Omega \vec k , \quad
  \vec I_{c,e} = \vec i , \quad
  \vec J_{c,e} = \vec j , \quad
  \vec K_{c,e} = \vec k ,
\end{split}
\label{eq.yego}
\end{equation}
where $C_o=C'_s-C'_c$. We further define $A_o=A'_s-A'_c$ and $B_o=B'_s-B'_c$. We then expand
$(\inertia_o)^{-1}$ in Taylor series up to the second order in $\vec y - \vec y_e$. This is
sufficient to get the equations of motion of the linearised system. We verify that $\vec y_e$
(Eq.~\ref{eq.yego}) actually is a solution of $\vec \nabla_{\vec y} H_\mathrm{go}(\vec y_e) = \sum_i
\mu_i \vec \nabla_{\vec y}C_\mathrm{go}^i(\vec y_e)$ where the Lagrange multipliers are
\begin{equation}
\begin{split}
& \mu_0 = 0 , \quad
  \mu_1 = 3A_{c}^{o}\sigma^0_{xx} + u_{xx}, \quad
  \mu_2 = 3B_{c}^{o}\sigma^0_{yy} + u_{yy}, \\
& \mu_3 = 3C_{c}^{o}\sigma^0_{zz} + (C_c+C'_c)\Omega^2 + u_{zz}, \quad
  \mu_4 = \mu_5 = \mu_6 = 0 , \quad
  \mu_7 = 3A_{s}^{o}\sigma^0_{xx} + u_{xx}, \\
& \mu_8 = 3B_{s}^{o}\sigma^0_{yy} + u_{yy}, \quad
  \mu_9 = 3C_{s}^{o}\sigma^0_{zz} + (C_s-C'_s)\Omega^2 + u_{zz}, \\
& \mu_{10} = \mu_{11} = \mu_{12} = 0 ,
\end{split}
\end{equation}
with 
\begin{equation}
A_{s}^{o} = A_s + A'_s , \qquad A_{c}^{o} = A_c-A'_c . 
\end{equation}
The same rules apply for $B_{s}^{o}$, $C_{s}^{o}$, $B_{c}^{o}$ and $C_{c}^{o}$.  Let us write the
matrix of the linearised problem in the variables
\begin{equation}
\begin{split}
&\delta \vec y^* = \Big(
 \delta \Pi_{s,z}, 
 \delta \Pi_{c,z}, 
 \delta I_{s,y}, 
 \delta I_{c,y} ,
 \delta \Pi_{s,x},               
 \delta I_{s,z},
 \delta \Pi_{c,x},
 \delta I_{c,z},
 \delta \Pi_{o,x},
 \delta \Pi_{s,y}, \\
&\delta J_{s,z},
 \delta \Pi_{c,y},
 \delta J_{c,z},
 \delta \Pi_{o,y},
 \delta \Pi_{o,z},
 \delta I_{s,x},
 \delta J_{s,y},
 \delta K_{s,z},
 \delta I_{s,y}+J_{s,x},
 \delta I_{c,x},
 \delta J_{c,y}, \\
&\delta K_{c,z},
 \delta I_{c,y}+J_{c,x},
 \delta I_{s,z}+K_{s,x},
 \delta J_{s,z}+K_{s,y},
 \delta I_{c,z}+K_{c,x},  
 \delta J_{c,z}              
+K_{c,y} \trans{\Big)} ,
\end{split}
\end{equation}
such that, with the driving perturbation, the system reads
\begin{equation}
\delta\dot{\vec y}^* - \mat A_\mathrm{go}^*(\vec y_e) \delta \vec y^* =
\delta{\vec z}_\mathrm{go}(t),
\qquad
\mat A_\mathrm{go}^*(\vec y_e) \equiv
   \left[\renewcommand{\arraystretch}{1.2}\begin{array}{cc|cc|c}
\mat 0 & -\mat A_\mathrm{go}^{12} & \mat 0 & \mat 0 & \cdot \\
\mat A_\mathrm{go}^{21} & \mat 0 & \mat 0 & \mat 0 & \cdot \\ \hline
\mat 0 & \mat 0 & \mat 0 & -\mat A_\mathrm{go}^{34} & \cdot \\
\mat 0 & \mat 0 & \mat A_\mathrm{go}^{43} & \mat 0 & \cdot \\ \hline
\mat 0 & \mat 0 & \mat 0 & \mat 0 & \mat 0
\end{array}\right] ,
\label{eq.lingo}
\end{equation}
with
\begin{eqnarray}
&&\mat A_\mathrm{go}^{12} = \begin{bmatrix}
 3(B_{s}^{o}-A_{s}^{o})(\kappa_1-\kappa_2) +U_{xy}  &  -U_{xy}\\
 -U_{xy} & 3(B_{c}^{o}-A_{c}^{o})(\kappa_1-\kappa_2)+U_{xy} \\
\end{bmatrix} , \ \qquad \\
&&\mat A_\mathrm{go}^{21} = \begin{bmatrix}
\Frac{1}{C_s} & 0 \\
0 & \Frac{1}{C_c}
\end{bmatrix} ,
\end{eqnarray}
and
\begin{eqnarray}
\label{eq.Ago34}
&&\mat A_\mathrm{go}^{34} = \begin{bmatrix}
-\Omega & \quad M^B_s+U_{yz}+F_{1,s}^B & 0 & -U_{yz}-F^B_3 &  F_{2,s}^B  \\[.2em]
 \Frac{1}{B_s} &  \Frac{C_s-B_s}{B_s}\Omega & 0 & 0 & 0 \\[.2em]
0 & -U_{yz}-F^B_3 & -\Omega & \quad M^B_c + U_{yz}+F^B_{1,c} & -F^B_{2,c} \\[.2em]
0 & 0 &  \Frac{1}{B_c} &   \Frac{C_c-B_c}{B_c}\Omega & 0 \\[0.7em]
0 & -F^B_{2,s} C_o\Omega & 0 & F^B_{2,c} C_o\Omega & -F^B_4 - \Omega
\end{bmatrix} , \quad \quad \\[1em]
\label{eq.Ago43}
&&\mat A_\mathrm{go}^{43} = \begin{bmatrix}
-\Omega & \quad M_s^A + U_{xz}+F_{1,s}^A & 0 & -U_{xz}-F_3^A & F_{2,s}^A \\[.2em]
\Frac{1}{A_s} & \Frac{C_s-A_s}{A_s}\Omega & 0 & 0 & 0 \\[.2em]
0 & -U_{xz}-F_3^A & - \Omega & \quad M_c^A + U_{xz}+F_{1,c}^A & -F^A_{2,c} \\[.2em]
0 & 0 & \Frac{1}{A_c} & \Frac{C_c-A_c}{A_c}\Omega & 0 \\[0.7em]
0 & -F_{2,s}^AC_o\Omega & 0 & F_{2,c}^AC_o\Omega & -F_4^A-\Omega
\end{bmatrix} .
\end{eqnarray}
In matrices $\mat A_\mathrm{go}^{34}$ and $\mat A_\mathrm{go}^{43}$
(Eqs.~\ref{eq.Ago34},\ref{eq.Ago43}), the interaction with the central planet is represented by the
terms
\begin{equation}
M_i^A = 3(C_i^o-A_i^o)\kappa_1 , \qquad
M_i^B = 3(C_i^o-B_i^o)\kappa_2 ,  \qquad i=s,c,
\end{equation}
the core/shell gravitational coupling through the ocean interface is given by
\begin{eqnarray}
&& U_{xy} \equiv u_{xy}+u_{yx}-u_{xx}-u_{yy} , \\
&& U_{xz} \equiv u_{xz}+u_{zx}-u_{xx}-u_{zz} , \\
&& U_{yz} \equiv u_{yz}+u_{zy}-u_{yy}-u_{zz} ,
\end{eqnarray}
From the expressions of $(u_{ab})_{a,b\in\{x,y,z\}}$ given in Eqs.~(\ref{eq.uxyza}-\ref{eq.uxyzb}),
we get
\begin{eqnarray}
\label{eq.Uxy}
&& U_{xy} = 2{\cal G}(B_c^o-A_c^o)(\rho_s (g_s-f_s) + (\rho_o-\rho_s)(g_o-f_o)) , \\
\label{eq.Uxz}                                                               
&& U_{xz} = 2{\cal G}(C_c^o-A_c^o)(\rho_s (h_s-f_s) + (\rho_o-\rho_s)(h_o-f_o)) , \\
\label{eq.Uyz}                                                               
&& U_{yz} = 2{\cal G}(C_c^o-B_c^o)(\rho_s (h_s-g_s) + (\rho_o-\rho_s)(h_o-g_o)) .
\end{eqnarray}
Finally, the remaining terms
\begin{align}
& F^K_{1,s} = \Frac{(C'_s-K'_c)(C'_s-K'_s)}{K_o}\Omega^2 , & \qquad K=A,B \\[0.7em]
& F^K_{1,c} = \Frac{(C'_c-K'_s)(C'_c-K'_c)}{K_o}\Omega^2 , & \qquad K=A,B \\[0.7em]
& F^K_{2,i} = \Frac{C'_i-K'_i}{K_o}\Omega , & \qquad i=s,c ,  \qquad K=A,B \\[0.7em]
& F^K_3 = \Frac{(C'_s-K'_s)(C'_c-K'_c)}{K_o}\Omega^2 , & \qquad K=A,B \\[0.7em]
& F^K_4 = \Frac{C_o-K_o}{K_o}\Omega , & \qquad K=A,B
\end{align}
are only present in the linearised system because of the rotation of the ocean. If the Casimir
$C_0(\vec y) = \vec \Pi_o\cdot \vec \Pi_o/2$ were set equal to zero, i.e., if the ocean were not
rotating with respect to the inertial frame, all $F^K_{1,i}$, $F^K_{2,i}$, $F^K_3$, and $F^K_4$,
with $K=A,B$ and $i=s,c$, would be nil. The same conclusion would hold if the kinetic energy of the
ocean $\trans{\vec \Pi}_o ( \inertia_o)^{-1} \vec \Pi_o/2$ were skipped from the Hamiltonian
$H_\mathrm{go}$. We thus interpret these terms as due to the centrifugal force felt by the ocean and
responsible for an additional pressure on the interfaces with the interior and the shell. In that
case -- i.e., if the kinetic energy of the ocean were dropped --, the ocean angular momentum $\vec
\Pi_o$ would be decoupled from the rest of the system.  A quick inspection of the last row and
column of the matrices $\mat A_\mathrm{go}^{34}$ and $\mat A_\mathrm{go}^{43}$ indeed shows that a
perturbation of $\vec \Pi_o$ would rotate at the eigenfrequency $\Omega$ with respect to the
laboratory frame, and would thus be fixed in the inertial frame.

We note that given the structure of the matrix $\mat A_\mathrm{go}^*$, the linearised system is
characterised by two libration frequencies in longitude and five frequencies associated with
libration in latitude and wobble.

For this problem, the Lyapunov function reads
\begin{equation}
\begin{split}
& N_\mathrm{go}(\vec y) =
\\ &
\frac{1}{2A_o}\left(\Pi_{o,x}-\frac{1}{2}(C'_c-A'_c)\Omega(I_{c,z}-K_{c,x})+\frac{1}{2}(C'_s-A'_s)\Omega(I_{s,z}-K_{s,x})\right)^2
\\ & +
\frac{1}{2B_o}\left(\Pi_{o,y}-\frac{1}{2}(C'_c-B'_c)\Omega(J_{c,z}-K_{c,y})+\frac{1}{2}(C'_s-B'_s)\Omega(J_{s,z}-K_{s,y})\right)^2
\\ & +
\frac{1}{2A_c}\left(\Pi_{c,x}+\frac{1}{2}(C_c-A_c)\Omega(I_{c,z}-K_{c,x})\right)^2
\\ & +
\frac{1}{2B_c}\left(\Pi_{c,y}+\frac{1}{2}(C_c-B_c)\Omega(J_{c,z}-K_{c,y})\right)^2
+ \frac{1}{2C_c}\Pi_{c,z}^2
\\ & +
\frac{1}{2A_s}\left(\Pi_{s,x}+\frac{1}{2}(C_s-A_s)\Omega(I_{s,z}-K_{s,x})\right)^2
\\ & +
\frac{1}{2B_s}\left(\Pi_{s,y}+\frac{1}{2}(C_s-B_s)\Omega(J_{s,z}-K_{s,y})\right)^2
+ \frac{1}{2C_s}\Pi_{s,z}^2
\\ &
+\frac{U_{xy}}{4}\left((I_{c,y}-J_{c,x})-(I_{s,y}-J_{s,x})\right)^2
+\frac{U_{xz}}{4}\left((I_{c,z}-K_{c,x})-(I_{s,z}-K_{s,x})\right)^2
\\ &
+\frac{U_{yz}}{4}\left((J_{c,z}-K_{c,y})-(J_{s,z}-K_{s,y})\right)^2
\\ &
+\frac{n_1^s}{2} (J_{s,x}-I_{s,y})^2
+\frac{n_2^s}{2} (I_{s,z}-K_{s,x})^2
+\frac{n_3^s}{2} (J_{s,z}-k_{s,y})^2 
\\ &
+\frac{n_1^c}{2} (J_{c,x}-I_{c,y})^2
+\frac{n_2^c}{2} (I_{c,z}-K_{c,x})^2
+\frac{n_3^c}{2} (J_{c,z}-k_{c,y})^2 ,
\end{split}
\end{equation}
with
\begin{equation}
\begin{split}
&
n_1^* = \frac{3}{4}(B_*^o-A_*^o)(\kappa_1-\kappa_2) , \quad
n_2^* = \frac{1}{4}(C_*^o-A_*^o)(\Omega^2+3\kappa_1) , \quad
\\ &
n_3^* = \frac{1}{4}(C_*^o-B_*^o)(\Omega^2+3\kappa_2) ,
\end{split}
\end{equation}
and where $*=s,c$. We deduce that the system is nonlinearly stable if the following conditions are
met
\begin{equation}
U_{xy}>0 ,\qquad
U_{xz}>0 ,\qquad
U_{yz}>0 ,\qquad
C_*^o > B_*^o > A_*^o  \qquad\mathrm{with}\quad
*=s,c .
\label{eq.cond1}
\end{equation} 
Using the expressions of $U_{xy}$, $U_{xz}$, and $U_{yz}$ (Eqs.~\ref{eq.Uxy}-\ref{eq.Uyz}) expanded
at first order in the equatorial and polar flatness, the conditions (\ref{eq.cond1}) are equivalent
to
\begin{equation}
\left\{
\begin{aligned}
&
\rho_s\frac{a_s-c_s}{a_s} + (\rho_o-\rho_s)\frac{a_o-c_o}{a_o} >
\rho_s\frac{a_s-b_s}{a_s} + (\rho_o-\rho_s)\frac{a_o-b_o}{a_o} > 0 ,
\\ &
C_*^o > B_*^o > A_*^o , \qquad *=s,c .
\end{aligned}
\right.
\label{eq.stab}
\end{equation}

Finally, as in the previous section, to get the forced solution, we decompose the driving excitation
$\delta \vec z_\mathrm{go}(t)$ as $(\delta \vec z_\mathrm{go}^1, \delta \vec z_\mathrm{go}^2, \vec
0)$, with
\begin{equation}
\delta\vec z_\mathrm{go}^1(t) = \begin{pmatrix}
3(B_s^o-A_s^o)\sigma^1_{xy} \\ 3(B_c^o-A_c^o)\sigma^1_{xy} \\ 0 \\ 0
\end{pmatrix} ,\qquad
\delta\vec z_\mathrm{go}^2(t) = \begin{pmatrix}
3(C_s^o-B_s^o)\sigma^1_{yz} \\
0 \\
3(C_c^o-B_c^o)\sigma^1_{yz} \\
0 \\ 0 \\
-3(C_s^o-A_s^o)\sigma^1_{xz} \\
0 \\
-3(C_c^o-A_c^o)\sigma^1_{xz} \\
0 \\ 0
\end{pmatrix} .
\end{equation}

\section{Application}
\label{sec.application}

\subsection{Io's libration modes}
\label{sec.io}
Io, one of the Galilean satellite of Jupiter, is assumed to have a liquid core \citep{Anderson96}.
Its rotation motion has already been studied within the Poincar\'e-Hough paradigm using a
Hamiltonian formalism \citep{Henrard08}. This analysis has then been extended using the same method
in \citep{Noyelles13, Noyelles14}. Although the approach in {\em ibid.} is Hamiltonian, it differs
from that described in Sect.~\ref{sec.liquid-core} which is expressed in non-canonical variables.
Here, we revisit the problem with the aim of validating our method and, more specifically, the
quasi-spherical approximation (Sect.~\ref{sec.fluid-approximation}).

\taborbio

\tabfreqio

The orbital and physical parameters of Io, which are summarised in Tab.~\ref{tab.orbIo}, are taken
from \citep{Noyelles13,Noyelles14}%
\footnote{Here and throughout the paper, we follow the IAU recommendations which state that the
symbol for a Julian year is ``\Year''. Hence, radian per year is written
``rad/a''.}. The eigenfrequencies $\omega_{u}$, $\omega_{v}$, $\omega_{w}$ and $\omega_{z}$ are
directly computed from the matrix $\mat A^*_\mathrm{fc}$ (Eqs.~\ref{eq.A1bis},\ref{eq.A2bis}) for
the Poincar\'e-Hough model (Sect.~\ref{sec.poincare-hough}), and from the matrix $\mat
A^*_\mathrm{fc'}$ (Eqs.~\ref{eq.A1ter},\ref{eq.A2ter}) for the quasi-spherical approximation
(Sect.~\ref{sec.fluid-approximation}).  Hereafter, the two models are referred to as ``model
$\mathrm{fc}$'' and ``model~$\mathrm{fc'}$'', respectively. The eigenfrequencies are then converted
into periods for a direct comparison with \citep{Noyelles14}. The correspondence between the
eigenperiods of {\em ibid.} and the eigenfrequencies of this work is
\begin{equation}
T_u = \frac{2\pi}{\omega_u} ,\qquad
T_v = \frac{2\pi}{\omega_v-\Omega} ,\qquad
T_w = \frac{2\pi}{\omega_w} ,\qquad
T_z = \frac{2\pi}{\omega_z} .
\label{eq.Tuvwz}
\end{equation}

The results are gathered in Tab.~\ref{tab.freqIo}.  We observe a good match between model
$\mathrm{fc}$ and that of \citep{Noyelles14} for $T_u$, $T_w$, and $T_z$ with a maximal error of
about 0.2\%.  There is a larger discrepancy between the two approaches in the case of $T_v$ with a
deviation of almost 6\%, but this eigenmode is more sensitive due to the small denominator
$\omega_v-\Omega$ (Eq.~\ref{eq.Tuvwz}). It is also very sensitive to the polar flattening of
the core \citep{Noyelles12}. Nevertheless, the agreement is satisfactory given that the methods to
compute the eigenperiods in both studies are very different. The eigenfrequencies given by models
$\mathrm{fc}$ and $\mathrm{fc'}$ are also very close to each other. Once again, the largest
discrepancy occurs for $T_v$, but here it does not exceed 0.5\%.  We thus conclude that the
quasi-spherical approximation is justified.

Figure~\ref{fig.em} represents the trajectories of the principal axes $\vec I$, $\vec J$, and $\vec
K$ in the laboratory frame $(\vec i, \vec j, \vec k)$ while the system stands in each of the
eigenmodes. The corresponding eigenfrequencies are recalled below each subfigure.  We recognise the
libration motions of a rigid satellite which the name of the eigenmodes have been taken from.  In
\citep{Henrard08} and in \citep{Noyelles13, Noyelles14}, the eigenmode associated with
$\omega_z$ is referred to as the free libration of the core.  Nevertheless, given the strong
similarity between the motions associated with $\omega_v$ and $\omega_z$, we chose to
attribute the same name ``libration in latitude'' for both of them. Furthermore, from the
observation of the surface only it is hardly possible to distinguish one from the other. Actually,
the distinction between the two modes lies in the relative position of $\vec \Pi_c$ and $\vec \Pi$,
as shown in Fig.~\ref{fig.cistrans}. When the satellite is in the eigenmode associated with
$\omega_v$, the two vectors are on the same side from the origin, while in the eigenmode of
frequency $\omega_z$ they are on opposite side. 

\figeigenmodes

\figcistrans

\subsection{Titan's equilibrium obliquity}
\label{sec.titan}

In this section, we analyse the rotation of Titan orbiting Saturn.  Several hints suggest that this
satellite holds a global ocean under its surface \citep[][and references therein]{Coyette16}. Among
these clues, an important one for our purpose is Titan's ``high'' obliquity of $0.32\deg$ which
could not be explained if the satellite were solid \citep{Bills11}. Nevertheless, a discrepancy
still persists between the observations and the expected obliquity associated with the
Cassini state, the latter remaining below $0.15\deg$ for a large class of interior models
\citep[e.g.,][]{Baland11}.  Therefore, it has been proposed that Titan's current obliquity is
amplified by a resonance with one of the remaining orbital forcing frequencies \citep{Baland11,
Noyelles14b}. 

In his abstract, \citet{Henrard08} wrote about Io that ``{\em{}the addition of a degree of freedom
(the spin of the core) with a frequency close to the orbital frequency multiplies the possibility of
resonances}''. In the case of Titan, we also have an additional degree of freedom in comparison to
the previous studies quoted above.
We thus expect our model to be able to tilt Titan's axis more
easily.

The orbital elements of Titan are taken from the ephemeris TASS1.6 \citep{Vienne95}. From the full
solution, we only retain the keplerian motion and the nodal precession of the orbit with respect to
the Laplace plane\footnote{Here, we define Titan's Laplace plane as the plane whose
orientation is given by the constant part of the inclination solution of TASS1.6.
\citep{Vienne95}}. These parameters are summarised in Tab.~\ref{tab.orbTitan}.  Regarding Titan
internal structure, we select two models proposed by \citet{Fortes12}, hereafter referred to as
model F1 and F2. They assume a global ocean with extreme densities equal to 1023\,kg/m$^{3}$ and
1281\,kg/m$^{3}$, respectively. In model F1, the ocean is a mixture of water and methanol, while in
model F2, the ocean is made of water and ammonia. Parameters of these interior models are summarised
in Tab.~\ref{tab.physTitan}. In both models, the average density is 1881\,kg/m$^3$ and the mean
moment of inertia $I/(mR^2)$ remains within the errorbars provided by \citet{Iess12}. The equatorial
flattening $\zeta$ is obtained by integration of Clairaut's equation \citep{Clairaut43} assuming an
hydrostatic equilibrium \citep[same as][]{Richard14}. The boundary semi-axes at volumetric mean
radius $R$ between two layers are given by \citep[e.g.,][]{Rambaux13}
\begin{equation}
a = R\left(1+\frac{7}{9}\zeta\right), \qquad
b = R\left(1-\frac{2}{9}\zeta\right), \qquad
c = R\left(1-\frac{5}{9}\zeta\right).
\end{equation}
The values of the derived parameters involved in the Hamiltonian $H_\mathrm{go}(\vec y)$
(Eq.~\ref{eq.Hgo}) are listed in Tab.~\ref{tab.derivTitan}.

\taborbtitan

\tabphystitan

\tabderivtitan

\tabeigenfreq

The eigenfrequencies computed for the two interior models F1 and F2 are shown in
Tab.~\ref{tab.eigenfreq}. For each model, we assume either a rotating or a static ocean with respect
to the inertial frame (see Sect.~\ref{sec.ocean}). For reference, we also provide the
eigenfrequencies assuming a fully rigid satellite.  To interpret these eigenfrequencies, the
associated trajectories of the vectors $(\vec I_c, \vec J_c, \vec K_c)$ and $(\vec I_s, \vec J_s,
\vec K_s)$ are displayed in Fig.~\ref{fig.libTitan}. We recognise librations in longitude at
$\omega_{u1}$ and $\omega_{u2}$, librations in latitude at $\omega_{v1}$, $\omega_{v2}$ and
$\omega_{v3}$, and wobbles at $\omega_{w1}$ and $\omega_{w2}$. From Tab.~\ref{tab.eigenfreq}, we
observe that each eigenmode has a specific range of frequencies.  Libration frequencies in latitude
are close to the mean motion $\Omega \approx 143.9240$\,rad/\Year. Frequencies of libration in
longitude are between 2 and 8\,rad/\Year, and the wobble is the slowest motion with frequencies
ranging between 0.01 and 0.2\,rad/\Year.

\figlibtitan

The condition for Titan to have a significant (shell) obliquity is that one of the libration
frequencies in latitude gets close to the excitation frequency of the perturbation
$\sigma^1_{xz}(t)$ (Eq.~\ref{eq.sig1}), namely, $\omega^1_{xz} = \Omega - \dot \Phi \approx
143.9330$\,rad/\Year. In the case of a rigid satellite there is no lever arm. The libration
frequency only depends on the total moments of inertia which are constrained by observations. This
frequency, equal to $143.9582$\,rad/\Year, leads to an obliquity of $0.113\deg$ which is about one
third of the actual value $\varepsilon_\mathrm{obs} = 0.32\deg$. 

\figpositiveobliq
\tabobliq

When the ocean is taken into account, the system has three distinct frequencies of libration in
latitude which can potentially be in resonance with the orbital precession rate. It should
nevertheless be stressed that when the rotation of the ocean is set to zero, the frequency
$\omega_{w3}$ in Tab.~\ref{tab.eigenfreq} is just the mean motion $\Omega$ which is not involved in
the tilting of the shell axis.  Titan's obliquities $\varepsilon$ computed with the different models
are gathered in Tab.~\ref{tab.obliq}. Note that we allow the obliquity to be negative as explained
in Fig.~\ref{fig.po}. As expected, within the ``static ocean'' hypothesis the ocean is not affected
by the perturbation $\sigma^1_{xz}$. Its obliquity is $\varepsilon_o = -i$, meaning that $\vec
\Pi_o$ remains aligned with the Laplace pole $\vec k$ which is the third axis of our laboratory
reference frame. The last two eigenfrequencies $\omega_{v1}$ and $\omega_{v2}$ are further away from
$\omega^1_{xz}$ than $\omega_{v3}$. They only produce a shell obliquity of $\varepsilon_s \approx
0.06\deg$ which is much lower than the observed one. Furthermore, this result does not
significantly vary from model F1 to model F2.

If the rotation of the ocean is set equal to the mean rotation of the satellite, $\omega_{w3}$ is
the eigenfrequency responsible for the tilt of Titan's shell spin pole. With the two models F1 and
F2 considered here, the results are still very low: $\varepsilon_s = 0.004\deg$ with model F1 and
$\varepsilon_s = 0.108\deg$ with model F2. However, the two values vary by a factor 27. A
modification of Titan's interior is thus more likely to produce the observed obliquity if the
rotation of the ocean is taken into account.

\figevolobliq

To illustrate this statement, we generate a series of interior models of Titan based on the model
F1. To simulate inhomogeneities in the shell, we slightly modify the equatorial flattening $\zeta_s$
of the surface from $11.890\times10^{-5}$ to the hydrostatic value $12.068\times10^{-5}$ given in
Tab.~\ref{tab.physTitan}. These numbers should be compared to the equatorial flattenings
computed with the two models provided by \citet{Iess12}, i.e., $11.911\times10^{-5}$ (SOL2) and
$12.005\times10^{-5}$ (SOL1a). To keep the global moments of inertia constant, the equatorial
flattening of all the other layers are refitted using Clairaut's equation. It has been checked that
all these models are nonlinearly stable according to the condition Eq.~(\ref{eq.stab}).
Figure~\ref{fig.obliq} displays the evolution of the libration frequencies in latitude $\omega_{v2}$
and $\omega_{v3}$ as a function of the surface equatorial flattening $\zeta_s$. When the rotation of
the ocean is considered (left plots), $\omega_{v3}$ varies sufficiently to cross the resonant
frequency $\omega^1_{xz}$ at $\zeta_s\approx11.97\times10^{-5}$ where, in the linear approximation,
the shell obliquity diverges. More interestingly, for $\zeta_s\approx11.94\times10^{-5}$, the driven
shell obliquity $\varepsilon_s$ is equal to the observed value $\varepsilon_\mathrm{obs}=0.32\deg$.
In comparison, when the ocean is assumed to be static (right plots of Fig.~\ref{fig.obliq}),
$\omega_{v3}$ remains strictly equal to $\Omega$ and $\omega_{v2}$ barely evolves. As a consequence,
the equilibrium shell obliquity remains practically constant close to $0.062\deg$.

\section{Conclusion}
\label{sec.conclusion}
This paper provides a general method for analysing the rotation dynamics of a rigid body with a
fluid internal layer. The study is performed in a non-canonical Hamiltonian formalism well adapted
to systems near relative equilibria such as synchronous satellites in a Cassini state. The Poisson
structure of the non-canonical Hamiltonian is here obtained by a Legendre transformation of the
corresponding Lagrangian written using Poincar\'e's formalism which makes use of the properties of
the Lie group acting on the configuration space.

With this approach, we have been able to treat the case of a satellite with a liquid core or with a
global underneath ocean in the exact same manner as that of a rigid satellite. All the difficulty is
in the calculation of the Lagrangian function -- and more specifically, of the kinetic energy of the
fluid layer -- in terms of generalised coordinates and Lie velocities. For a satellite with a liquid
core, Poincar\'e introduced the concept of a fluid simple motion which cannot be rigorously
transposed to a satellite with an ocean. Nevertheless, at first order this fluid layer behaves like
a rigid body for which the kinetic energy is known. Tests on a satellite with a liquid core,
assuming Io's physical and orbital parameters, have shown that the errors induced by this
approximation do not exceed 0.5\% on the eigenfrequencies.

The analysis of a hollow satellite with a fluid core leads to a four degree of freedom dynamical
model. The linearised problem in the vicinity of the synchronous equilibrium state is thus
characterised by four eigenmodes. These are a libration in longitude, a wobble and two librations in
latitude. To this solved problem, we have provided an analytical expression of the linearised
equations written in terms of intuitive variables, namely the components of the angular momenta and
of the base frame vectors. We also have clearly identified the fourth eigenmode as a libration in
latitude.

The rotation dynamics of a satellite with a global subsurface ocean is governed by seven eigenmodes
associated with the seven degrees of freedom of the problem, six of which being equally
shared by the interior and the outer shell and the last one being brought by the ocean. Near the
synchronous equilibrium state, these eigenmodes are identified as two librations in longitude, two
wobbles and three librations in latitude.
The amplitude of the third libration in latitude would only vanish if the ocean were static
with respect to the inertial frame.

Our study has been motivated by Titan's obliquity measured by the Cassini-Huygens mission. Thus far,
dynamical models struggle to explain its high value under the hydrostatic shape hypothesis
suggested by the ratio of its Stokes coefficients $J_2/C_{22}\approx 10/3$. Here, we show that the
rotation of the ocean makes the dynamical model much more sensitive to small perturbations of the
interior model than when the ocean is assumed static. As an example, starting from a body in perfect
hydrostatic equilibrium, we slightly modified the equatorial flattening of the shell by about 1\% of
the nominal value. This was enough to bring the obliquity of the Cassini state even beyond the
radiometric value with the seven degree of freedom model while the same quantity computed with the
static ocean hypothesis remained practically constant scarcely reaching a 0.1\% increase.

This work is intended to demonstrate the capability of the seven degree of freedom dynamical model
to explain the observed high obliquity of Titan. The problem has therefore been intentionally
simplified. Tidal deformations, atmospheric torques, and all orbital perturbations but the main
precession relative to the Laplace plane have been discarded.  These additions would be required for
an exhaustive search of the interior models compatible with the measurements made by the
Cassini-Huygens mission: the rotation state, the gravity field coefficients, the shape, the tidal
Love number, and the electric field.  But this is beyond the scope of the present paper and shall be
discussed elsewhere.

\begin{acknowledgements}
We thank Beno{\^i}t Noyelles and the anonymous referee for their constructive comments. G.B. also
thank Philippe Robutel for the useful conversations on the theoretical parts of this work and
Rose-Marie Baland and Marie Yseboodt for our instructive discussion on this problem during the 2017
DDA meeting in London.
\end{acknowledgements}

\bibliographystyle{cemda}         
\bibliography{titan}   

\end{document}